\documentclass[fleqn,usenatbib]{mnras}

\usepackage{newtxtext,newtxmath}
\usepackage[T1]{fontenc}
\usepackage{ae,aecompl}

\usepackage{graphicx}	% Including figure files
\usepackage{amsmath}	% Advanced maths commands
\usepackage[usenames,dvipsnames]{color}
\usepackage[]{color}

\usepackage{graphicx}	% Including figure files
\usepackage{amsmath}	% Advanced maths commands

\usepackage{hyperref} 
\usepackage{setspace}
\title[UV background and star formation in simulated dwarfs]{The cosmic UV background and the beginning and end of star formation in simulated field dwarf galaxies}

\author[M. Pereira-Wilson et al.]{
Matthew Pereira-Wilson$^{1\thanks{E-mail: mhpereir@gmail.com}}$,
Julio F. Navarro$^{1}$, 
Alejandro Ben\'\i tez-Llambay$^{2}$, and
 Isabel Santos-Santos$^{3}$ \\
$^{1}$Department of Physics and Astronomy, University of Victoria, Victoria, BC V8P 5C2, Canada \\
$^{3}$Dipartimento di Fisica G. Occhialini, Universit\`a degli Studi di Milano Bicocca, Piazza della Scienza, 3 I-20126 Milano MI, Italy\\
$^{2}$Institute for Computational Cosmology, Department of Physics, University of Durham, South Road, Durham DH1 3LE, UK}

\defcitealias{BLF21}{BLF21}
\defcitealias{BLF20}{BLF20}
\defcitealias{BL17}{BL17}

\date{Accepted XXX. Received YYY; in original form ZZZ}

\pubyear{2021}

\begin{document}
\label{firstpage}
\pagerange{\pageref{firstpage}--\pageref{lastpage}}
\maketitle

\begin{abstract} We use the APOSTLE cosmological simulations to examine the role of the cosmic UV background in regulating star formation (SF) in low-mass $\Lambda$CDM halos. In agreement with earlier work, we find that after reionization SF proceeds mainly in halos whose mass exceeds a redshift-dependent ``critical'' mass, $M_{\rm crit}$, set by the structure of the halos and by the thermal pressure of UV-heated gas. $M_{\rm crit}$ increases from $\sim 10^{8}\, M_\odot$ at $z\sim 10$ to $M_{\rm crit}\sim 10^{9.7}\, M_\odot$ at $z=0$, roughly following the average mass growth of halos in that mass range. This implies that halos well above or below critical at present have remained so since early times. Halos of luminous dwarfs today were already above-critical and star-forming at high redshift, explaining naturally the ubiquitous presence of ancient stellar populations in dwarfs, regardless of luminosity.  The SF history of systems close to the critical boundary is more complex. SF may cease or reignite in dwarfs whose host halo falls below or climbs above the critical boundary, suggesting an attractive explanation for the episodic nature of SF in some dwarfs. Also, some subcritical halos today may have been above critical in the past; these systems should at present make up a sizable population of faint field dwarfs lacking ongoing star formation. Although few such galaxies are currently known, the discovery of this population would provide strong support for our results.  Our work indicates that, rather than stellar feedback, it is the ionizing UV background and mass accretion history what regulates SF in the faintest dwarfs.  \end{abstract}


\section{Introduction}
\label{SecIntro}

The Lambda Cold Dark Matter ($\Lambda$CDM) paradigm for structure formation makes a number of well-defined and robust predictions for the formation and evolution of dark matter halos. In particular, $\Lambda$CDM predicts a power-law halo mass function at the low-mass end that is much steeper than the galaxy stellar mass function at the faint end.  This discrepancy places strong constraints on the galaxy mass-halo mass relation at the faint end, and is usually explained by arguing for a steady decrease in galaxy formation ``efficiency'' with decreasing halo mass, so that, effectively, no luminous galaxies form below some ``threshold'' halo mass \citep[see, e.g., the excellent review of][and references therein]{Bullock_2017}.

The origin of this threshold is usually ascribed to the effects of the ionizing UV background~\citep{Rees_1986,Ikeuchi_1986,Efstathiou_1992} which heats the gas of the Universe after cosmic reionization, preventing its collapse and subsequent transformation into stars in the shallow potential wells of low-mass halos \citep{Quinn_1996,Thoul_1996,Navarro_1997a}.

Although the importance of the UV background has long been recognized as a mainstay of $\Lambda$CDM dwarf galaxy formation models \citep[see; e.g., early work by][]{Couchman_1986, Chiba_1994, Bullock_2000, Benson_2002, Somerville_2002}, there is still disagreement about how it translates in practice into regulating the star formation history in low-mass halos and whether it actually implies the actual presence of a minimum ``threshold'' halo mass for galaxy formation \citep[see; e.g.,][]{Wheeler_2019,Nadler_2020}.

%These disagreements  extend to our understanding of the processes that regulate the wide variety of SFH in dwarfs, which vary from systems that formed all their stars at early times, to systems where star formation is long-lasting and still ongoing, to systems punctuated by episodes of star formation separated by relatively quiescent periods \citep[][]{Grebel_2004, Ricotti_2005}. 

%In particular, questions regarding how cosmic reionization affects dwarfs, or how energetic feedback from evolving stars may affect the aforementioned halo mass ``threshold'' for galaxy formation, are still heavily debated.

For example, it would be natural to expect in this scenario that reionization should have had a defining effect on the formation of dwarf galaxies and that these effects may have left a recognizable imprint in the star formation history (SFH) of dwarfs.
Since cosmic reionization is widely thought to have happened rather early and abruptly, early models suggested that it should have left a similarly abrupt signature in the SFHs of the faintest dwarfs, namely a sharp truncation in their star formation at the time of reionization \citep{Gnedin_1997,Ricotti_2002b}.

However, exquisite panoramic data from HST, coupled with the latest stellar population synthesis models, have revealed that even non-star-forming dwarfs dominated by old stellar populations seem to have had protracted star formation activity extending well past the epoch of reionization \citep{Weisz_2011, Weisz_2014, Weisz_2014b, Gallart_2015, Skillman_2017}.  Indeed, only a few extremely faint satellites of the Milky Way (MW) seem to have a stellar population consisting solely or mainly of stars formed before reionization \citep{Brown_2014}. How does then reionization affect star formation in a dwarf, and how do those effects compare to others, driven perhaps by environment or feedback? Does the UV background play a defining role in ending or modulating the star formation history of a dwarf?

A similar set of questions are posed by other properties of dwarf galaxies, such as the ubiquitous presence of ancient stellar populations \citep[][and references therein]{Weisz_2014, Skillman_2017}. These suggest that the onset of star formation happened very early in all dwarfs, regardless of their present-day luminosity. This is somewhat surprising in the context of a ``threshold'' for galaxy formation since, presumably, systems of different mass would reach this threshold at different times.

In addition, if the ionizing UV background was partly responsible for curbing star formation in some dwarfs, it may have also played a role in the wide variety of dwarf SFHs. These vary from systems that formed all their stars at early times, to systems where star formation is long-lasting and still ongoing, to systems punctuated by episodes of star formation separated by relatively quiescent periods \citep[][]{Mateo_1998, Grebel_2004, Tolstoy_2009, Simon_2019}. 

Dwarf galaxy properties are also known to vary greatly with the environment. For example, only two of the satellites of the MW are currently forming stars \citep[the Magellanic Clouds; the rest are all quiescent dwarf spheroidals, or dSphs, see; e.g.,][]{McConnachie_2012}, but most ``field'' dwarfs outside the virial radius of a more massive host appear to be actively forming stars at present \citep{Geha_2012}. This clear environmental dependence may reflect, however, a subtle mass dependence; indeed, most satellites of the MW are intrinsically much fainter than the majority of field dwarfs studied so far. How do the effects of the ionizing UV background depend on galaxy mass?

Early work on this topic focused mainly on the statistics of the dwarf population, and aimed at establishing the characteristic ``filtering'' mass of halos whose baryonic content is severely affected by reionization \citep[e.g.,][]{Gnedin_2000,Okamoto_2008}, rather than on the effects of reionization on the star formation history of individual dwarf systems. This is now changing, however, as cosmological hydrodynamical simulations start to tackle the low-mass halo regime \citep[][]{Simpson_2013,Shen_2014, BL15,Jeon_2017,Fitts_2017,Maccio_2017,Wheeler_2019,Wright_2019, Munshi_2019,Applebaum_2021,Rey_2022,Gutcke_2022}, and as theoretical modelling considers the evolving thermodynamics of photoionized gas in a hierarchically evolving population of cold dark matter halos.

The halo mass function and its redshift evolution are now understood. It is also widely accepted that the mass profile of virialized halos is well approximated at all times and for all masses by the Navarro-Frenk-White profile \citep[hereafter, NFW,][]{Navarro_1996,Navarro_1997b}, with parameters that are well specified for $\Lambda$CDM at all redshifts \citep[see; e.g., the ``mass-concentration relation'' of][and references therein]{Ludlow_2016}.

Before star formation begins in earnest, the properties of (primordial) gas in such halos are also well understood. This is particularly true after reionization, when the interplay between gas heating by UV photons and gas cooling from collisional excitation of H and He, lead to a tight link between gas density and temperature. This ``equation of state'' may be used to infer the gas density and temperature profile in halos of arbitrary mass (assuming hydrostatic equilibrium), enabling simple assessments regarding which systems can form stars and when. 

In particular, UV-heated gas in equilibrium in an NFW halo follows a well-specified density profile as a function of halo mass, concentration, and redshift\footnote{Assuming that systems are dominated gravitationally by dark matter.}. This profile, coupled with a suitably specified criterion (e.g., total gas content or a threshold central density), can be used to infer which halos may be susceptible to begin forming stars. This procedure translates into a well-defined, redshift-dependent characteristic ``critical'' halo mass, $M_{\rm crit}(z)$, above which star formation begins \citep[][hereafter BLF20]{BLF20}.

Halos below critical contain tenuous photo-heated gas supported by their own pressure, and they remain ``dark'' provided they are sub-critical at all times. We shall refer to such systems hereafter as RELHICs (short for ``REionization-Limited HI Clouds'') following \citet{BL17} (hereafter BL17).

Within the~\citetalias{BLF20} framework, $M_{\rm crit}(z)$ can be combined with the mass accretion history of an individual halo to ascertain when star formation begins \citep[][hereafter BLF21]{BLF21}. Importantly, the same star formation criterion also implies that star formation may cease should the halo become sub-critical at later times. This suggests that the effects of the ionizing UV background on systems close to the critical boundary may be variegated and spread over time, depending on the vagaries of each halo's individual mass accretion history.

Our main goal is to assess whether it is possible to understand the beginning and end of star formation in simulated faint dwarfs in terms of the previous framework. To this end, we use cosmological hydrodynamical simulations of the APOSTLE project \citep{Sawala_2016,Fattahi_2016}. These simulations use the EAGLE code \citep{Schaye_2015,Crain_2015} to evolve volumes tailored to resemble the Local Group and its surroundings. The simulations contain hundreds of low-mass halos, both isolated (``field'') systems and sub-halos embedded within the virial\footnote{We define the virial radius, $r_{200}$, of each halo as that enclosing a sphere of mean density equal to $200\times$ the critical density for closure, $\rho_{\rm crit}=3H(z)^2/8\pi G$, where $H(z)$ is Hubble's ``constant''. We denote values computed within or at the virial boundary with a ``200'' subscript.} boundaries of more massive systems. For simplicity, our analysis focuses on field systems only, since sub-halos are likely affected by environmental effects, such as tidal and ram-pressure stripping, which may overwhelm or confuse the effect of the ionizing UV background.

We also restrict much of our analysis to times after the epoch of reionization, which is assumed to be $z_{\rm reion}\approx 11.5$ in the simulations. As we discuss below, the onset of star formation before reionization is artificially curtailed in our simulations by the adoption of an effective polytropic equation of state (PEoS) affecting high-density gas. This PEoS is adopted in APOSTLE to prevent spurious fragmentation in star-forming regions, but it also imposes an effective ceiling on the gas density of early-collapsing clumps which delays the onset of star formation in many of them. Although this limitation may induce a short delay in the onset of star formation of some systems, it should not invalidate our main conclusions.
% JFN--The time at which the first stars forms in our systems is thus an upper limit that does not preclude qualitatively the results that follow.

 We organize the paper as follows. We begin with a brief description of the APOSTLE simulations in Sec.~\ref{SecNumSims} before describing the analytical ``critical'' mass modelling in Sec.~\ref{SecCritMass}. We contrast these analytic results with APOSTLE results in Sec.~\ref{SecRes} and conclude with a brief discussion and summary in Sec.~\ref{SecConc}.

 \section{The APOSTLE simulations}
 \label{SecNumSims}

The APOSTLE\footnote{"A Project Of Simulating The Local Environment"} simulation suite consists of 12 zoom-in cosmological hydrodynamical simulations. The simulated volumes were selected from the DOVE N-body simulation \citep{Jenkins_2013}, with the intent of reproducing the Local-Group environment. Each volume has a pair of halos with mass and kinematic properties similar to those of the Milky Way-Andromeda system \citep{Fattahi_2016}. In our analysis we use the five volumes evolved at the highest mass resolution, with gas and dark matter particle masses $m_{\text{gas}} = 1\times 10 ^{4}M_\odot$ and $m_{\text{dm}}=5\times 10^{4}M_\odot$, respectively, and a Plummer-equivalent gravitational softening length, $\epsilon = 134$ pc. APOSTLE adopts the WMAP-7 cosmological parameters \citep{Komatsu_2011}.

The APOSTLE simulations were performed using a modified version of P-GADGET3 code \citep{Springel_2005} developed for the EAGLE cosmological simulation \citep{Crain_2015, Schaye_2015}. The adjustable numerical parameters used in APOSTLE were the same as in the EAGLE reference runs. We briefly describe some aspects relevant to our analysis below, and refer the reader to the reference EAGLE papers for full details.

\subsection{Radiative cooling, UV-background, and cosmic reionization}

Radiative cooling and photoheating rates correspond to those of \cite{Wiersma_2009}, calculated using the code CLOUDY \citep{Ferland_1998}, under the assumption that the gas is dust-free, optically thin, and in ionization equilibrium. After the redshift of reionization, assumed to be $z_{\rm reion}=11.5$ in the simulations, the gas is exposed to the time-evolving, but spatially uniform~\cite{Haardt_2001} ionizing UV background. To account for the fact that the gas is not optically thin before reionization, an extra 2 eV per proton mass is added, which ensures the intergalactic gas is quickly ionized and heated to $\approx 10^4$K. For hydrogen, this is done at $z=11.5$, while for helium the extra energy is distributed in redshift with a Gaussian centred at $z=3.5$, of width $\Delta z=0.5$.

\subsection{Star formation}
\label{SecSF}

Like EAGLE, the APOSTLE simulations are not intended to resolve the multiphase ISM or cold molecular gas complexes ($T<<10^{4}$K). In order to prevent numerical instabilities on such small scales, the simulation imposes a minimum pressure floor on the gas, which takes the form of a ``polytropic equation of state'' (PEoS),  
\begin{equation}
P_{\text{EoS}}= P_0 \bigg(\frac{\rho_g}{\rho_0}\bigg)^{\Gamma},% \hspace{0.5cm} P_{\text{IGL}}<P_{\text{EoS}}
\label{EqPEoS}
\end{equation}
with $\Gamma=4/3$ and where $P_0=1.1 \rm ~g ~cm^{-1} ~s^{-2} $ and $\rho_0/m_p=0.1 \rm ~cm^{-3}$.
%In addition to the PEoS, the simulation also imposes a temperature floor of $8000$~K for gas densities $n_H>10^{-5} \rm ~cm^{-3}$, in order to further prevent metal-rich gas from forming dense clumps. 
In practice, this forces high-density gas to have a temperature that simply reflects the effective pressure of the unresolved ISM, and cannot be trusted for other physical considerations, such as calculating neutral hydrogen fractions in post-processing. 

Given the lack of modelling of cold molecular gas, the simulation allows star formation to proceed in gas particles whose density exceeds the threshold above which a cold phase may form. This is chosen to be $10 \text{ cm}^{-3}$ for primordial gas but allowed to decrease with increasing metallicity, $Z$, in enriched gas particles \citep{Schaye_2004};
\begin{equation}
    n_{\rm thr}(Z)=\text{min} \bigg[ 10^{-1}\text{ cm}^{-3}\bigg(\frac{Z}{0.002}\bigg)^{-0.64}, 10 \text{ cm}^{-3} \bigg].
\end{equation}
For the systems we focus on in this paper, the maximum threshold is the more relevant since it is the one applicable to primordial/low metallicity gas.

Finally, because gas density in the early universe was very high, a simple physical density threshold may allow star formation everywhere at very high redshifts. For this reason, an overdensity requirement is also imposed, with the density of gas particles having to exceed 57.7 times the cosmic mean for them to be eligible to turn into stars. This choice of overdensity requirement does not significantly affect the results \citep{Schaye_2015}, mainly due to the imposition of the PEoS, which prevents gas from reaching high densities in low-mass systems at high redshift.

\subsection{Stellar Feedback}

Stellar particles are treated as simple stellar populations (SSPs) with a \citet{Chabrier_2003} initial mass function (IMF) in the mass range $0.1$-$100M_\odot$. The energy feedback from SNIa implementation follows \citet{Dalla_2012}, where stellar particles release their feedback energy in a stochastic manner to individual gas particles nearby. The energy received by each gas particle is such that the particle increases its temperature by $\Delta T=10^{7.5}K$, with the probability that any gas particle be heated proportionally to the total amount of energy released by the SSP, which corresponds to the release of $10^{51}$ erg per supernova, and assumes that stars with masses $6$-$100\, M_\odot$ explode via this channel.

\subsection{Halo finder}

Substructures in the simulation are identified using the SUBFIND group finder \citep{Springel_2001, Dolag_2009}. Halos are first identified by running a friends-of-friends algorithm \citep[FoF;][]{Davis_1985} on the dark-matter particles, with a linking length 0.2 times the mean interparticle separation. Gas and stellar particles are then assigned to the same FoF group as their nearest dark-matter particle. SUBFIND, using all particles, then recovers gravitationally bound substructures within each FoF group, which we refer to as subhaloes. In this work, we only study the properties of the main (``central'') subhalo of each FoF halo. 

We use in our analysis all central halos found within a spherical volume of radius $3$ Mpc, centred on the barycenter of the two main halos in each volume. The barycenter is calculated for each snapshot, spanning the redshift range from $0$  to $20$. We restrict our analysis to halos with $M_{200}>10^{7}M_\odot$, or the equivalent of about $200$ dark matter particles. In practice, we shall see that no halos below $10^8\, M_\odot$ are able to form stars in these APOSTLE runs, so our low-mass limit does not preclude the results that follow, and our analysis concerns mainly halos resolved with an equivalent of at least $2000$ dark matter particles.

\section{Critical virial mass for the onset of star formation}
\label{SecCritMass}

\subsection{The BLF20 model}

The~\citetalias{BLF20} model establishes the ``critical'' virial mass needed for the onset of star formation in a halo. To calculate the ``critical'' mass, this model assumes that post-reionization, the density profile of gas in a CDM halo is such that the gas is in thermal equilibrium with the ionizing UV background, in hydrostatic equilibrium with the halo, and with an outer pressure  ($r\rightarrow\infty$) that corresponds to that of the intergalactic medium at the mean density. For simplicity, dark halos are assumed to be spherically symmetric and well approximated by NFW profiles. The criterion for the onset of star formation is based on the total gas content within the virial boundary of a halo, obtained by integrating the gas density profile. Beyond some redshift-dependent minimum mass, $M_{\rm crit}(z)$, the gas content exceeds $f_{\rm bar}\, M_{200}$, the total baryonic mass expected within $r_{200}$ according to the universal baryon fraction, $f_{\rm bar}=\Omega_{\rm bar}/\Omega_{\rm M}$. The theoretical total gas mass quickly diverges for masses greater than $M_{\rm crit}$, indicating that pressure alone cannot stop gas from flowing to the centre of a halo, where it should turn into stars.

Prior to reionization, the intergalactic medium is not pressurised beyond the virial radius of the dark halos, and one cannot use the same~\citetalias{BLF20} boundary condition to derive their gas density profile. However, one can choose a boundary condition such that the gas density at the virial radius is set so that the total enclosed gas-mass within $r_{200}$ equals the universal baryon fraction, i.e., $M_{\rm gas}(r<r_{200}) = f_{\rm bar} M_{200}$. The temperature of the gas inside the halos roughly corresponds to the virial temperature, ensuring the systems remain in equilibrium unless cooling becomes important. The~\citetalias{BLF20} model assumes that, prior to reioonization, star formation  proceeds predominantly in halos whose virial mass exceeds the atomic cooling limit, or $T_{200} = 7\times 10^3$~K. 
% JFN-This is the reason why the~\citetalias{BLF20} assumes that star formation proceeds predominantly in halos whose virial mass exceeds the atomic cooling limit, or $T_{200} = 7\times 10^3$~K.

\subsection{A density-threshold criterion for the critical mass}

We adopt here a different criterion to derive the ``critical'' mass for the onset of star formation based on the central (maximum) density of the gas rather than the total gas mass within the halo. This criterion is better attuned to the choices made in cosmological hydrodynamical simulations, which often rely on a minimum ``threshold'' gas density for star formation to proceed. We show below in Sec.~\ref{SecCompMPBL} that, after reionization,  this criterion returns results for $M_{\rm crit}(z)$ similar to those obtained with the~\citetalias{BLF20} model. Prior to reionization, however, its predictions differ from BLF20, mainly because of the adoption of  the PEoS at high densities (see Eq.~\ref{EqPEoS}), which leads to an artificial lower limit of $\sim 10^8\, M_\odot$ in the APOSTLE critical mass.

\begin{figure}
    \centering
    \includegraphics[width=0.49\textwidth]{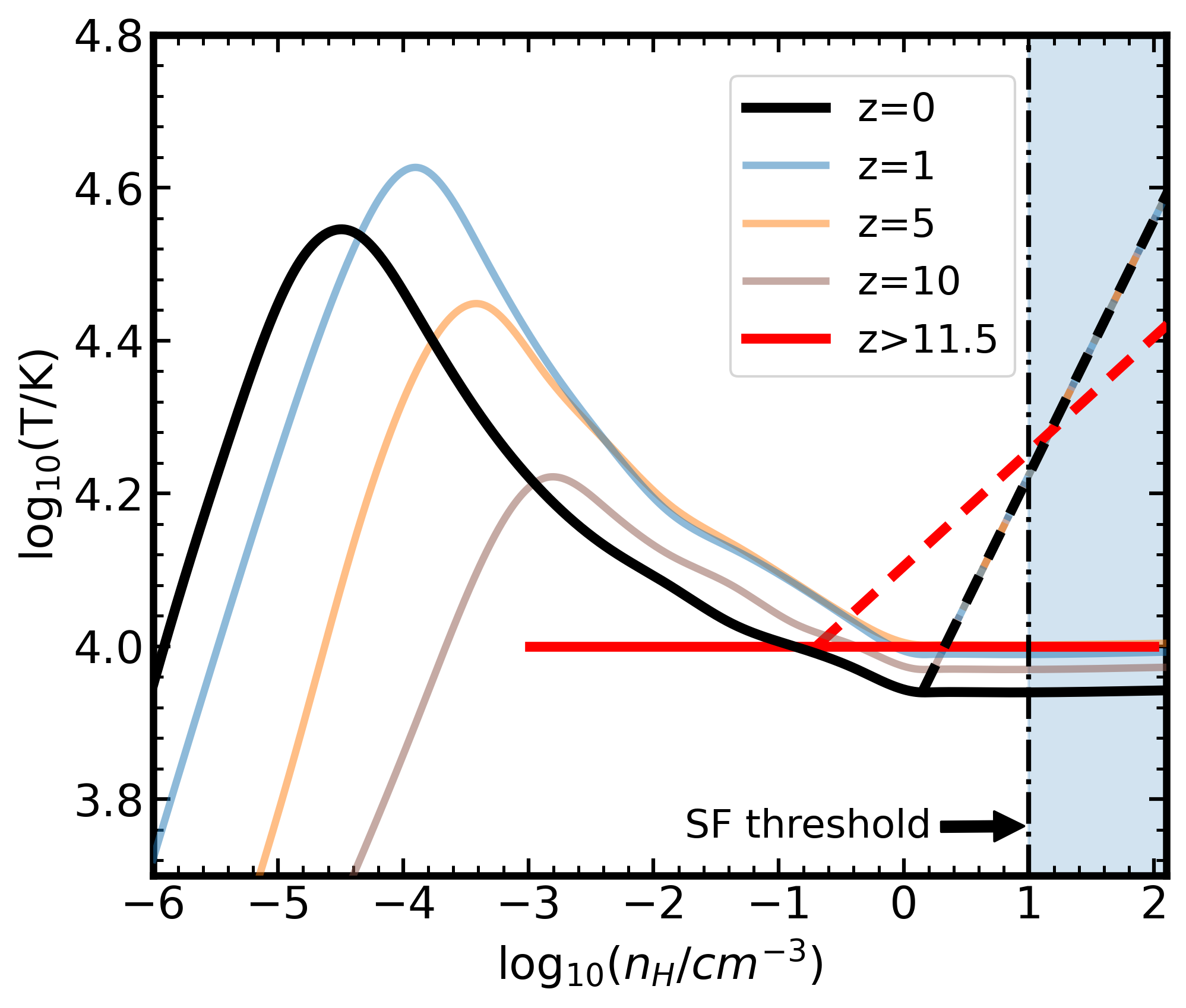}
    \caption{ Gas temperature-density relation assumed in our modeling at different redshifts. At high densities,  dashed lines indicate the polytropic equation of state (PEoS) adopted in the APOSTLE/EAGLE simulations. In that regime, solid lines assume that the gas is isothermal. Prior to reionization, our model assumes that, at low densities, the gas inside the virial radius of critical halos is isothermal at $10^4$~K (thick red line). Densities above the APOSTLE star-formation threshold ($n_H=10\, \text{cm}^{-3}$) are highlighted in light blue. }
    \label{fig:trho}
\end{figure}%

\subsubsection{Temperature-density relation}

The post-reionization temperature-density relation of photoionized gas is shown, at various redshifts, in Fig.~\ref{fig:trho}. As discussed by~\citetalias{BL17}, this relation is characterized by two different regimes. At low densities, the temperature rises steeply, as a result of photoheating. This rise tracks the loci where the photoheating timescale equals the age of the Universe at that redshift, and therefore it shifts to higher densities at earlier times.

The maximum of each curve corresponds to a density where the photoheating rise meets the loci where photoheating and radiative cooling timescales are comparable. At larger densities,  the $T$-$\rho$ curve drops from its maximum and approaches roughly $10^4$ K, the minimum temperature needed to collisionally excite the Ly-$\alpha$ transition \citep[see; e.g.,][and references therein]{Haehnelt_1996,Theuns_1998}.

The sharp upturn in temperature at high density (shown by dashed lines) is not physical and corresponds to the imposed pressure minimum of the PEoS, which applies only to high-density gas in the APOSTLE/EAGLE simulations (see Sec.~\ref{SecSF}). This PEoS is implemented as a single pressure-density relation, but manifests itself as two different $T$-$\rho$ relations, before and after $z_{\rm reion}$, because of the change in molecular weight that occurs at reionization.
As we shall see below, the adoption of a PEoS curtails (artificially) the ability of low-mass halos to form stars before reionization.

Before reionization, our model assumes that the gas can cool efficiently, and thus remains isothermal at $10^{4}$~K in halos able to form stars (thick red line in Fig.~\ref{fig:trho}). The isothermal assumption only applies within the virial radius; we highlight this by truncating the red horizontal line in  Fig.~\ref{fig:trho} at $n_H=10^{-3}~\rm cm^{-3}$.  Since atomic cooling is the main cooling mechanism for primordial gas in APOSTLE halos, this is an adequate assumption that may be used to derive the gas density profile of a halo at the critical boundary. In practice, we shall see below that the adoption  of a PEoS imposes a minimum mass for the onset of star formation well above that corresponding to a virial temperature of $10^{4}$~K, so the isothermal assumption used by the model to derive the critical mass before reionization seems well justified.

%JFN--in halos is isothermal at $10^{4}$~K within the virial radius (thick red line in , for densities not affected by the PEoS. Although this assumption is realistic only for halos more massive than the atomic cooling limit, for which gas can cool, it is in these halos where we expect the gas to become unstable and form stars. Therefore, the assumed temperature is adequate to derive the gas density profile, and therefore the ``critical'' mass, of halos that can harbour star formation prior to reionization. To highlight that the isothermal assumption does not apply outside the virial radius, the line shown in Fig.~\ref{fig:trho} is truncated at $n_H=10^{-3}~\rm cm^{-3}$. 

\subsubsection{Analytic gas density profiles}

\begin{figure}
    \centering
%    \begin{minipage}{0.49\textwidth}
    \includegraphics[width=0.49\textwidth]{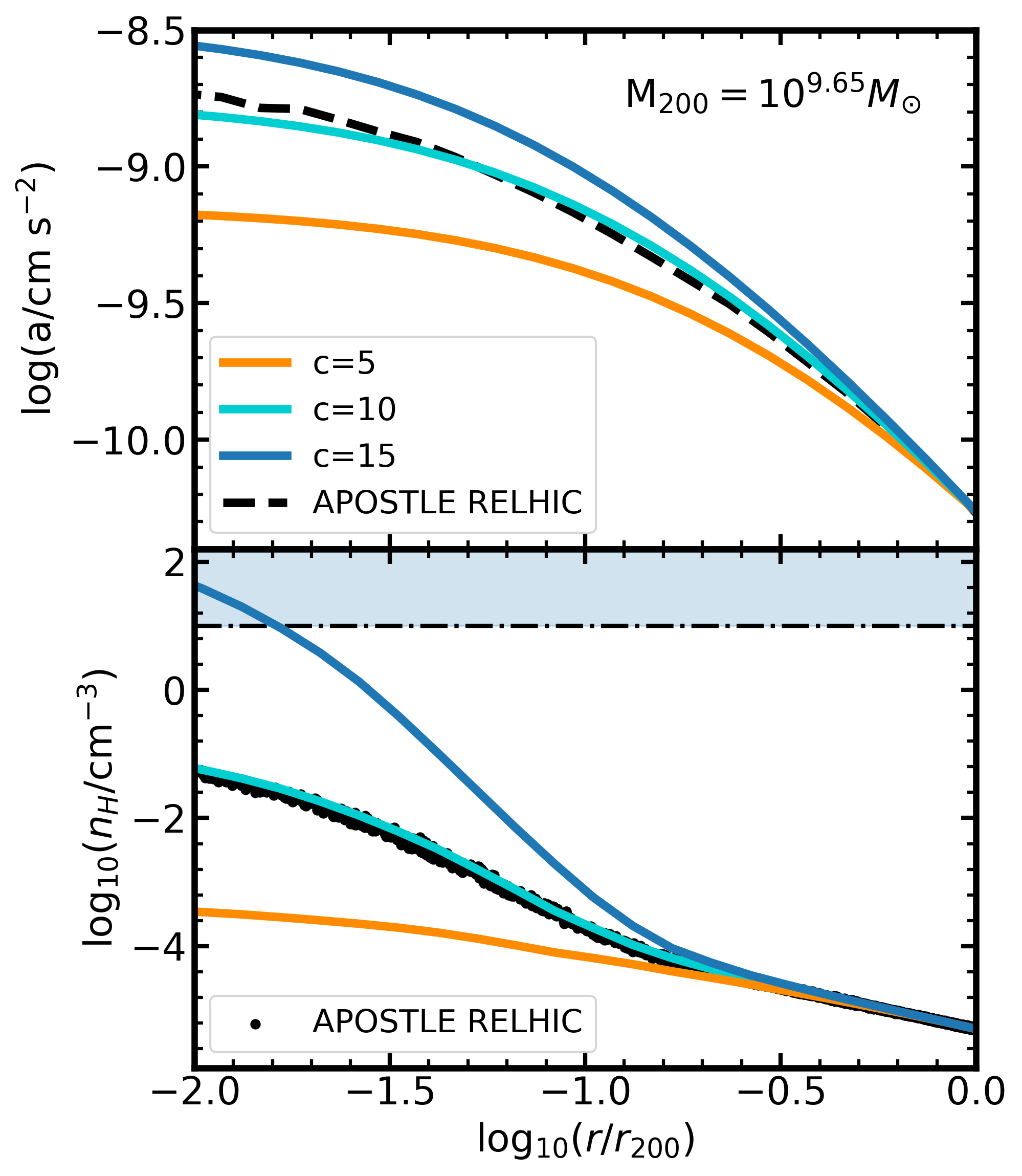}
    \caption{ {\it Top:} Acceleration profile ($a(r)=GM(r)/r^2$) of  $M_{200}=10^{9.65}M_\odot$ NFW halos of different concentration at $z=0$ (solid coloured lines).  The dashed black curve is the acceleration profile of an APOSTLE RELHIC of the same  virial mass. {\it Bottom:} Gas density profile of the RELHIC (black dots) as well as model density profiles computed assuming hydrostatic equilibrium; solid coloured lines). Densities above the APOSTLE star-formation threshold are highlighted in blue. }\vspace{0.2cm}
    \label{fig:rho_profile_1}
%    \end{minipage}
\end{figure}

Using the described $T$-$\rho$ relations, and following the procedure outlined by~\citetalias{BLF20} (see their Sec. 2), we may now compute the gas density profile of a ``dark'' (RELHIC) halo at any redshift, using the appropriate $T$-$\rho$ relation, as shown in Fig.~\ref{fig:trho}. We illustrate this in Fig.~\ref{fig:rho_profile_1}, where the dashed curve in the top panel shows the total gravitational acceleration profile, $a(r)=GM(r)/r^2$, of an APOSTLE RELHIC at $z=0$. The dots in the bottom panel indicate the gas profile of this same RELHIC. The NFW profiles of three halos with the same virial mass ($M_{200}=10^{9.65} M_\odot$), but different concentrations ($c=5$, $10$, and $15$), are also shown with thick coloured lines. For the appropriate concentration ($c\approx 10$), the analytic gas profile matches the density profile of the simulation data remarkably well\footnote{To best fit the data of this example RELHIC, we set a boundary condition equal to the RELHIC gas density at $r=r_{200}$, instead of the condition at infinity which we use in the critical mass modelling.} (see Fig.~5 in BL17 for a similar example).

The shaded region in the bottom panel of  Fig.~\ref{fig:rho_profile_1} indicates densities above the star formation threshold assumed for primordial gas in our simulations. At fixed halo mass, the resulting density profile is highly dependent on halo concentration; for $c=10$, the gas in the halo would remain in hydrostatic equilibrium without forming stars, whereas, for $c=15$, a halo of the same mass would begin to form stars at its centre. This exercise illustrates that it is not only the halo mass but also its concentration that determines which halos undergo star formation and host a luminous galaxy. We return to the impact of concentration on the value of the ``critical'' mass in Sec.~\ref{SecHaloConc}.

\begin{figure}
%    \begin{minipage}{0.49\textwidth}
    \includegraphics[width=0.49\textwidth]{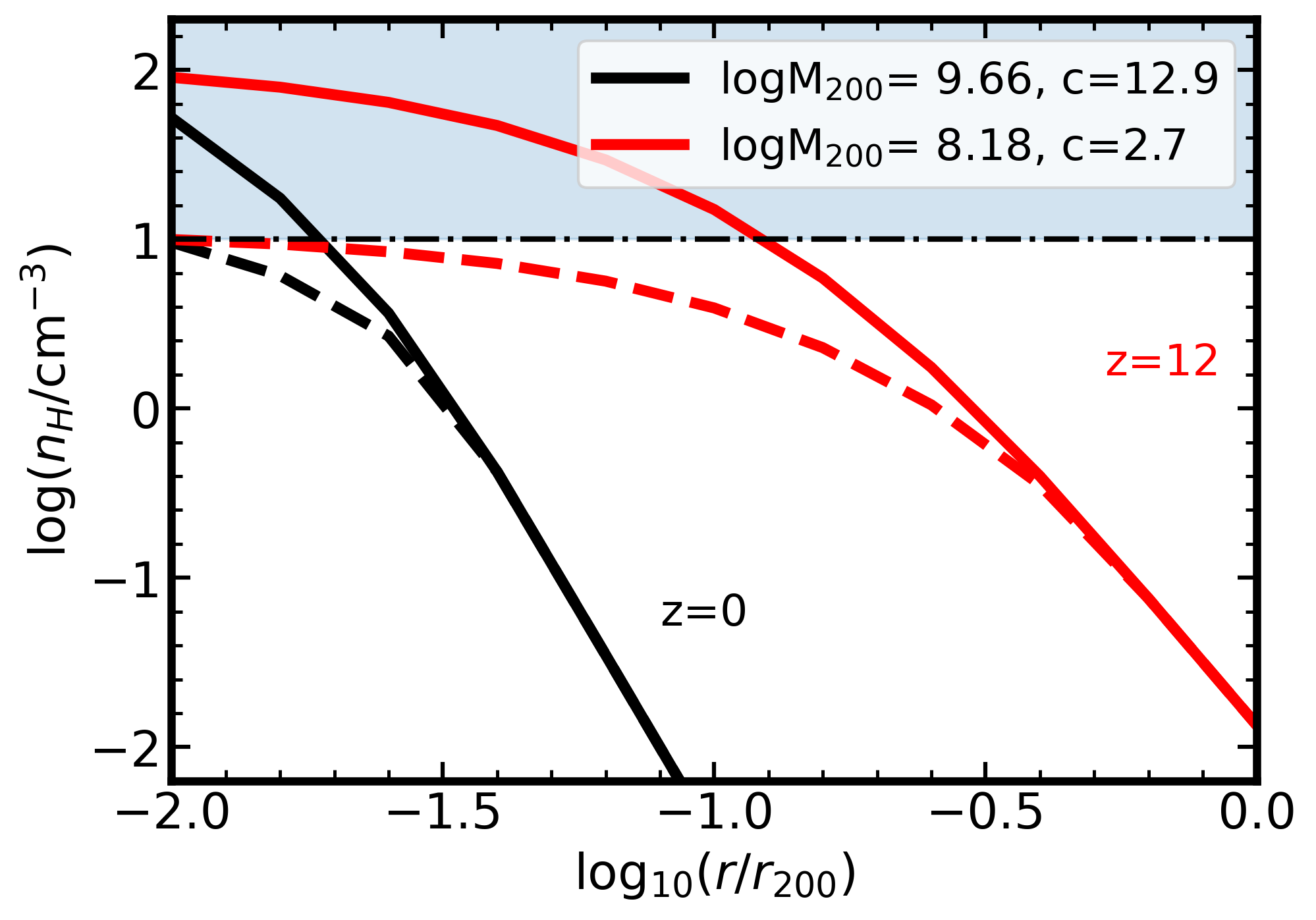}
\caption{ Model gas density profiles assuming that the gas remains isothermal at high densities (solid curves) or that it follows the APOSTLE/EAGLE polytropic equation of state (PEoS; dashed curves). Two NFW halos are shown, at $z=0$ and at $z=12$, with masses and concentrations as listed in the legend. Densities above the APOSTLE star-formation threshold are highlighted in blue.}
    \label{fig:rho_profile_2}
%    \end{minipage}
\end{figure}%

We examine next the role of the PEoS imposed on high-density gas on the onset of star formation in our simulation. Fig.~\ref{fig:rho_profile_2} shows the gas density profiles of two halos at two different redshifts, computed including the PEoS (dashed lines) or not (solid). The concentration of each halo corresponds to the average expected for their mass given the \citet{Ludlow_2016} mass-concentration-redshift relation. The adoption of a PEoS clearly depresses the central gas densities. This is true in particular prior to reionization where the PEoS renders most halos below $\sim 10^8\, M_\odot$ (artificially) ineligible for star formation. Indeed, as we shall see below, all APOSTLE halos that begin forming stars at $z>z_{\rm reion}$ exceed a virial mass of $\sim 10^8\, M_\odot$ (a mass that is larger than the atomic cooling limit), an artificial result likely arising from the PEoS.

\subsubsection{Halo concentration and central gas density}
\label{SecHaloConc}

\begin{figure}
    \centering
    \includegraphics[width=0.49\textwidth]{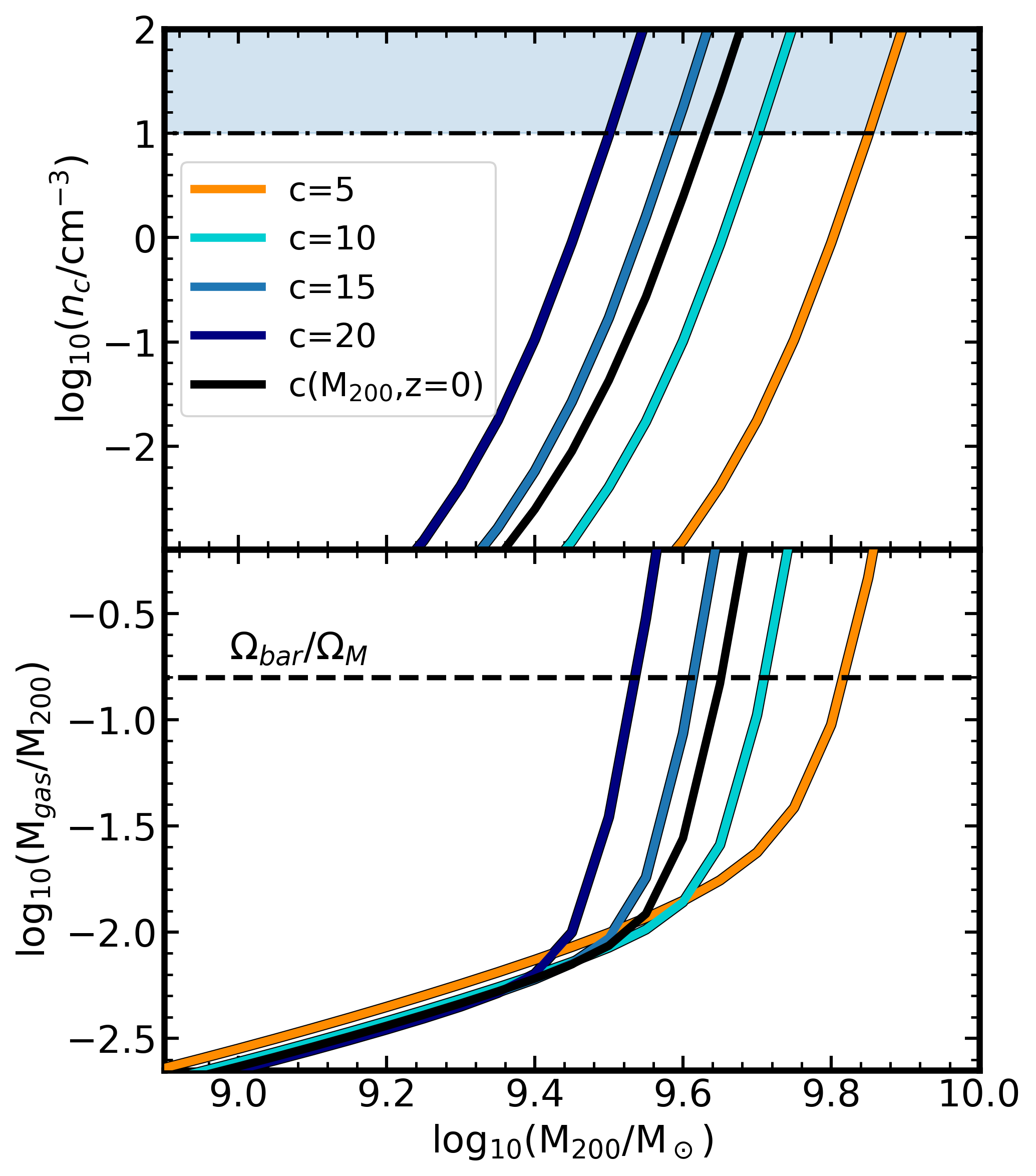}
    \caption{ {\it Top:} Central gas densities, $n_c$ (defined as $n_H$ at $r=0.01\, r_{200}$), calculated assuming hydrostatic equilibrium and the $T$-$\rho$ relation without PEoS (solid curves in Fig.~\ref{fig:trho}).  Four NFW profiles with different concentrations are shown, including one (shown in black) that follows the average concentration of LCDM halos, which in this mass range is $c\sim 13$ \citep{Ludlow_2016}. {\it Bottom:} Total gas mass within the virial radius for halos shown in the top panel.}
    \label{fig:mgas_c}
\end{figure}

To explore and isolate the effects of concentration on the gas profile, we adopt the $T$-$\rho$ relation without a PEoS in this section. In Fig.~\ref{fig:mgas_c} we summarise the effects of concentration on gas properties. The top panel shows, as a function of halo mass at $z=0$, the impact on the central gas density (defined as $n_c$, or the density at $r=0.01\, r_{200}$) of varying the average halo concentration about the value, $c\approx 13$, expected for $\Lambda$CDM \citep{Ludlow_2016}. Average-concentration halos (solid black curve) are expected to become eligible for star formation for virial masses exceeding $10^{9.63}M_\odot$, but this boundary varies somewhat for lower or higher-than-average concentration halos.

However, the variation of the ``critical'' halo mass is not large, only about a factor of $\sim 2$ for concentrations between $5$ and $20$. We thus conclude that although halo concentration plays a role in determining which halos remain ``dark'' or host galaxies, it appears to be secondary compared to the role of halo mass. Indeed, we show this in the bottom panel of Fig.~\ref{fig:mgas_c}, which is analogous to the top, but displays the total gas mass expected within the virial radius and how it varies with halo concentration. For average-concentration halos (solid black curve), the total gas mass equals the expected gas content of the halo at $M_{200}\sim 10^{9.65} M_\odot$ and diverges rapidly at higher masses: gas in such halos is unable to stay in hydrostatic equilibrium and would collapse to the centre and trigger the onset of star formation. This is indeed the rationale for the ``critical mass'' for star formation advocated by~\citetalias{BLF20}.

Comparing the top and bottom panels of Fig.~\ref{fig:mgas_c} shows that defining the ``critical mass'' either by total gas content or central gas density gives similar results (within a factor of $\sim 2$; note that none of these curves includes the effects of the EAGLE PEoS). This provides further evidence for the robustness of the concept of critical mass.

\begin{figure}
    \centering
    \includegraphics[width=0.49\textwidth]{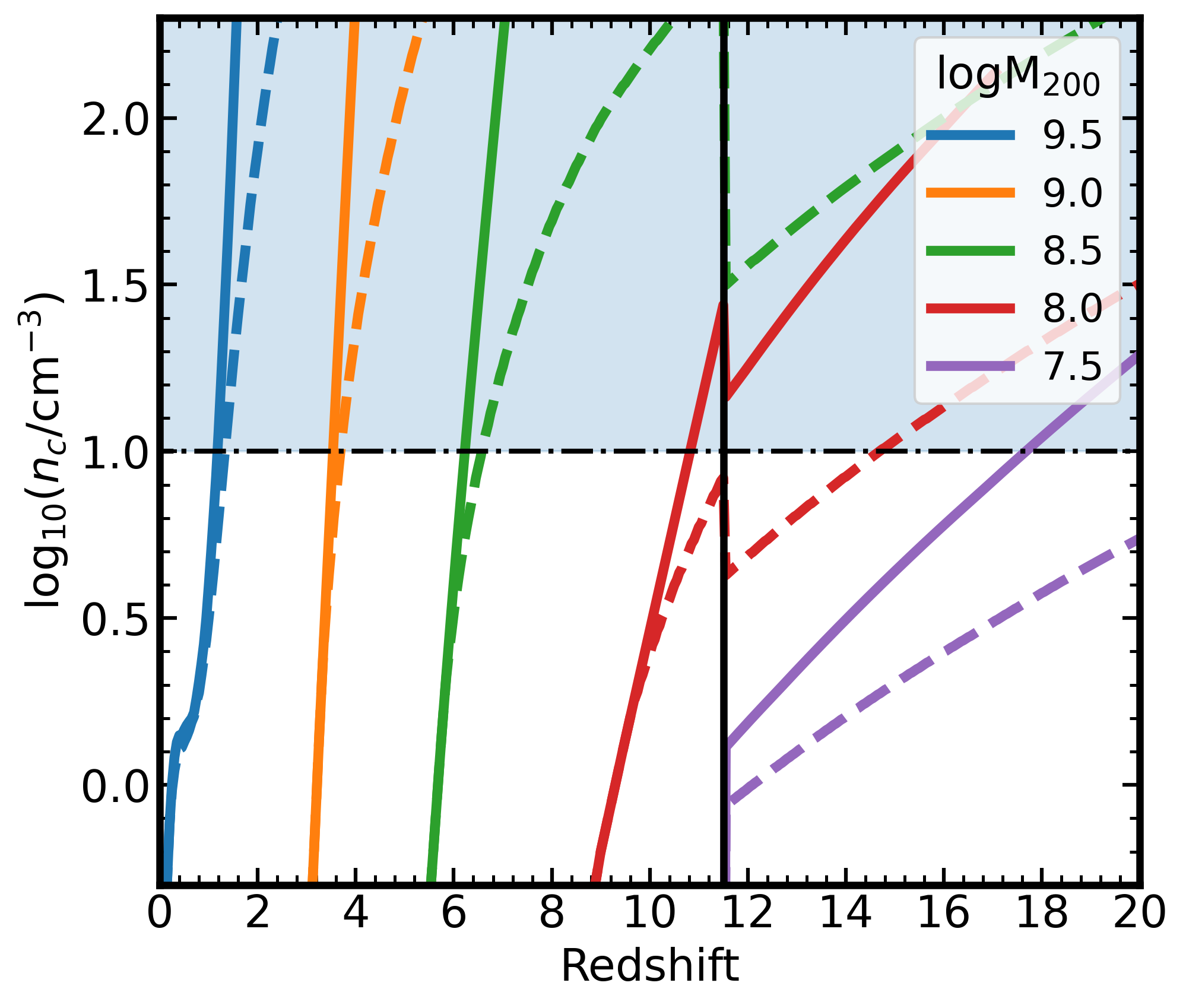}
    \caption{Central gas densities, $n_c$, computed assuming hydrostatic equilibrium and that the gas is isothermal at high densities (solid curves) or that it follows the APOSTLE/EAGLE PEoS (dashed curves). Each curve corresponds to halos of different virial mass, as listed in the legend. Each of these halos becomes ``critical'' at the redshift where they cross the star-formation threshold. The reionization redshift is shown by the black vertical line. Densities above the star-formation threshold of APOSTLE are highlighted by the shaded blue region.}
    \label{fig:rhoc}
\end{figure}

Finally, we explore in Fig.~\ref{fig:rhoc} how the central gas density varies as a function of redshift for halos of different masses, as labelled. Solid curves use the $T$-$\rho$ relations of Fig.~\ref{fig:trho} and assume average concentrations for each halo. The dashed curves correspond to models that include the PEoS modification at high gas densities implemented in EAGLE/APOSTLE. The intersection of each curve with the central density threshold of $n_{c}=10 \text{ cm}^{-3}$ corresponds to the redshift at which that mass equals $M_{\rm crit}$,  the "critical" mass in APOSTLE.

Each of these curves assumes the concentration-mass-redshift dependence of \citet{Ludlow_2016}. The critical mass decreases systematically with redshift, driven primarily by the increase in gas density and the evolution of the $T$-$\rho$ relation. The critical mass is also sensitive to the PEoS assumption, particularly at high redshifts where the difference between the two sets of lines grows.

\subsection{Critical mass model comparison: \citetalias{BLF20} vs APOSTLE}
\label{SecCompMPBL}

We are now ready to compute a critical mass as a function of redshift that can be compared directly with the results of the APOSTLE simulation. This is shown by the black dashed curve in Fig.~\ref{fig:mcrit}, which traces the halo mass that hosts a system with $n_c=10$ cm$^{-3}$, the APOSTLE primordial gas density threshold for star formation. The black dashed curve assumes average concentrations from the \citet{Ludlow_2016} $c(M,z)$ relation and the $T$-$\rho$ relations (including the effects of EAGLE's PEoS) shown in Fig.~\ref{fig:trho}.

We compare the APOSTLE critical mass with the critical mass from~\citetalias{BLF20}, shown by the thick magenta curve in Fig.~\ref{fig:mcrit}. We calculate this curve using the total gas mass criterion illustrated in the bottom panel of Fig.~\ref{fig:mgas_c}. Following ~\citetalias{BLF20}, the model adopts, for simplicity, a constant concentration of $c=10$, and that the critical mass is approximated by the atomic cooling mass, with virial temperature, $T_{200}=7\times10^3$~K, prior to reionization.

The most notable change between the black and magenta lines is that the jump to lower $M_{\rm crit}$ at $z>z_{\rm reion}$ clearly seen in the~\citetalias{BLF20} model is affected when introducing the PEoS. In aggregate, these changes reduce substantially the central gas densities that the gas may reach in halos near the critical boundary before reionization. Because of this, we expect only APOSTLE halos that exceed $10^{7.7}$-$10^{8.2}\, M_\odot$ to be able to start forming stars before reionization ($z_{\rm reion} = 11.5$). Note that this is higher than either the critical boundary expected either from the H-cooling limit (set at $7000$~K; magenta curve) or from $H_2$-cooling, indicated by the blue dotted line \citep{Tegmark_1997}.

\section{Results}
\label{SecRes}

\subsection{The onset of star formation in APOSTLE halos}

%\begin{figure}
%    \centering
%    \includegraphics[width=0.49\textwidth]{images/tlast_m200.png}
%    \caption{Distribution of quenching times as a function of present day $M_{200}$, shown as the lookback time at which the last star (youngest star) formed within a halo. Galaxies that formed their last star $<0.5$ Gyrs from today are considered star-forming (blue), otherwise are quiescent (red). The present day critical mass is shown as a vertical dashed line and the epoch of reionization as a horizontal solid line. }
%    \label{fig:tlast_m200}
%\end{figure}

We begin by analysing how well the critical mass model discussed in the previous section describes the onset of star formation in the APOSTLE simulation. This is shown in Fig.~\ref{fig:mcrit_2}, where we plot the virial mass of a halo at the time it forms its first star. Individual systems are shown with squares, coloured by their concentration, computed from the ratio between the maximum circular velocity, $V_{\rm max}$, and the circular velocity at the virial radius, $V_{200}$, assuming NFW profiles. The critical mass curves from~\citetalias{BLF20} (solid magenta) and from our simulation model (black) are also shown.

Despite the simplicity of the critical mass model, it appears to capture well the main trends highlighted in Fig.~\ref{fig:mcrit_2}. In particular, it is clear that the minimum mass needed to ignite star formation increases steadily with decreasing redshift, and that the boundary is well approximated by the critical mass. The jump of the critical mass at the redshift of reionization in our simulations seems to differ from that predicted by the~\citetalias{BLF20} model. As discussed above, this is expected from the introduction of the artificial PEoS in APOSTLE. Indeed, the black curve model, which includes the PEoS, is in much better agreement with the simulation data before reionization. We conclude that APOSTLE systematically underestimates the redshift at which early-collapsing halos may start forming stars before reionization.
%JFN--, and possibly also the total number of luminous systems formed.

%It may also under-predict the total number of dwarf galaxies, particularly because it suppresses star-formation in halos that should form stars in halos above the atomic cooling limit prior to reionization, but which don't grow above APOSTLE critical mass line at any time.
%The exact number of galaxies in this range should be rather limited, and can be calculated as a correction to the APOSTLE halo occupancy function. 

A few other points are worth noting in Fig.~\ref{fig:mcrit_2}. One is that, at fixed redshift, the scatter in halo mass at the time of first star formation correlates with halo concentration: the higher the concentration the lower the mass needed to trigger star formation, as expected from our discussion in Sec.~\ref{SecHaloConc} (see Fig.~\ref{fig:rho_profile_1}).

\begin{figure}
    \centering
    \includegraphics[width=0.48\textwidth]{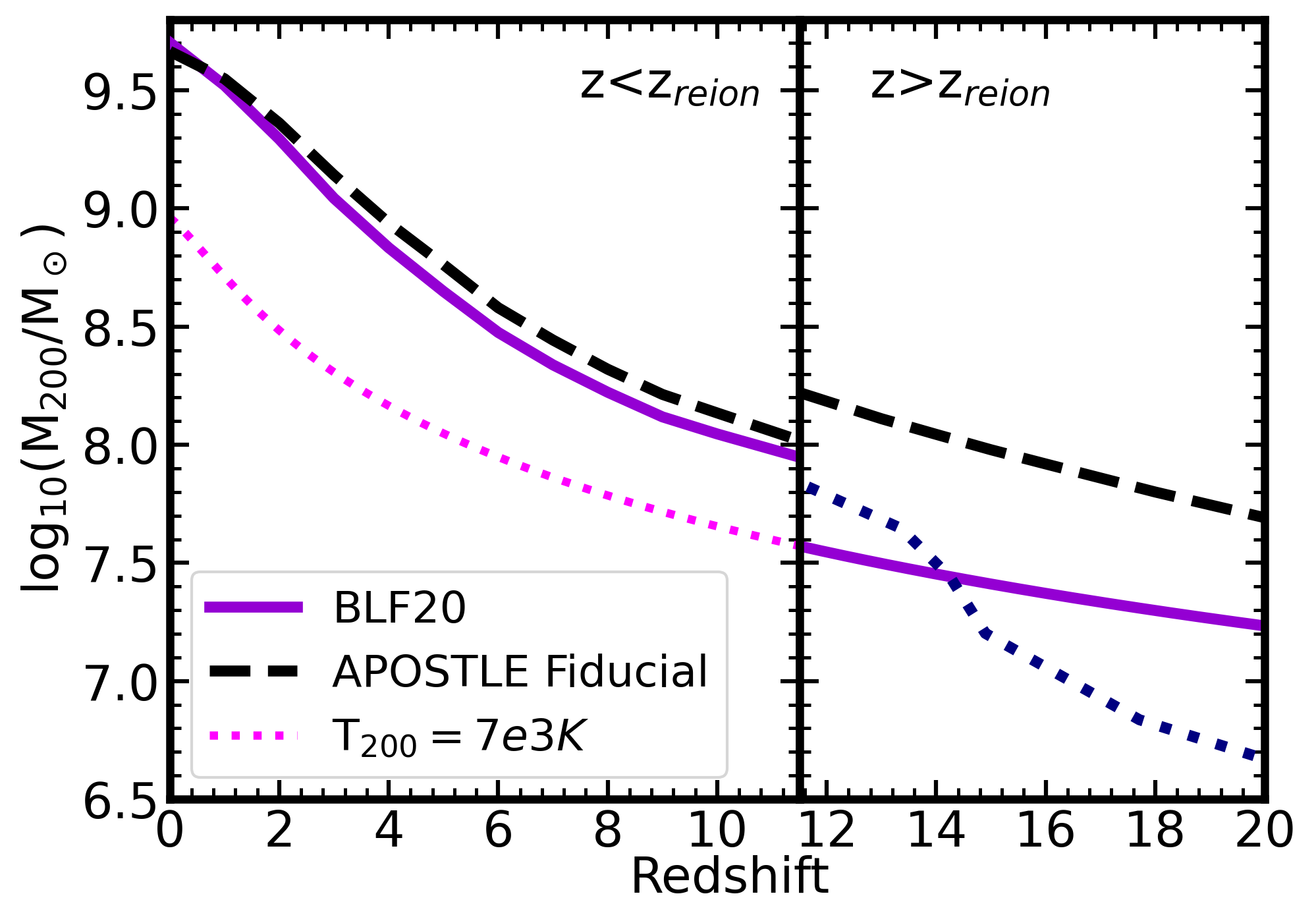}
    \caption{ Critical mass curves as a function of redshift: solid purple curve uses the \citet{BLF20} definition. Our fiducial  APOSTLE model is shown in dashed-black, calculated with the $T$-$\rho$ relations shown in Fig.~\ref{fig:trho}, including the PEoS. For reference, the magenta dotted line show the virial mass for a fixed $T_{200}=7000$~K, whereas the blue dotted line tracks, prior to reionization, the $H_2$-cooling critical mass from \citet{Tegmark_1997}. }
\label{fig:mcrit}
\end{figure}%

\begin{figure*}
    \centering
    \includegraphics[width=0.98\textwidth]{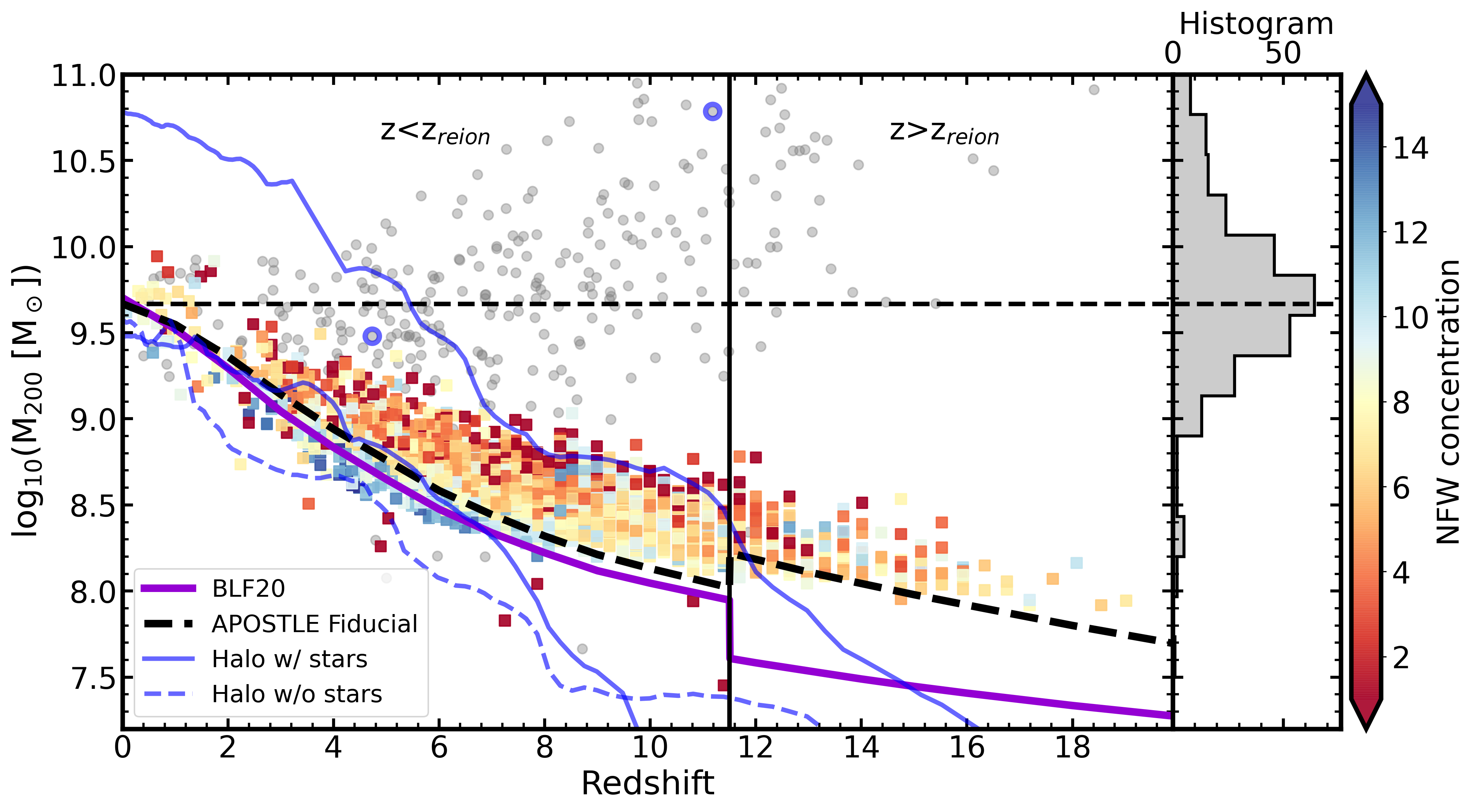}
    \caption{Squared symbols show the virial mass of APOSTLE halos at the time when their first star is formed, colored by concentration. The magenta curve is the BLF20 critical mass; the dashed black curve is the APOSTLE fiducial model from Fig.~\ref{fig:mcrit}. Grey circles indicate the redshift of formation of the oldest star as a function of virial mass for APOSTLE centrals with $M_{200}< 10^{11}\, M_\odot$ at $z=0$. The blue lines show three example halo mass assembly histories. Two of the halos host luminous galaxies at $z=0$ (solid blue);  the other hosts a star-less RELHIC (dashed blue). The time at which the two luminous halos began star-forming are highlighted by a blue-circle. The histogram in the right panel shows the distribution of $M_{200}$ for the grey data points, with a clear peak near the critical mass at $z=0$.}
    \label{fig:mcrit_2}
  \end{figure*}%
  
\begin{figure}
    \centering
    \includegraphics[width=0.48\textwidth]{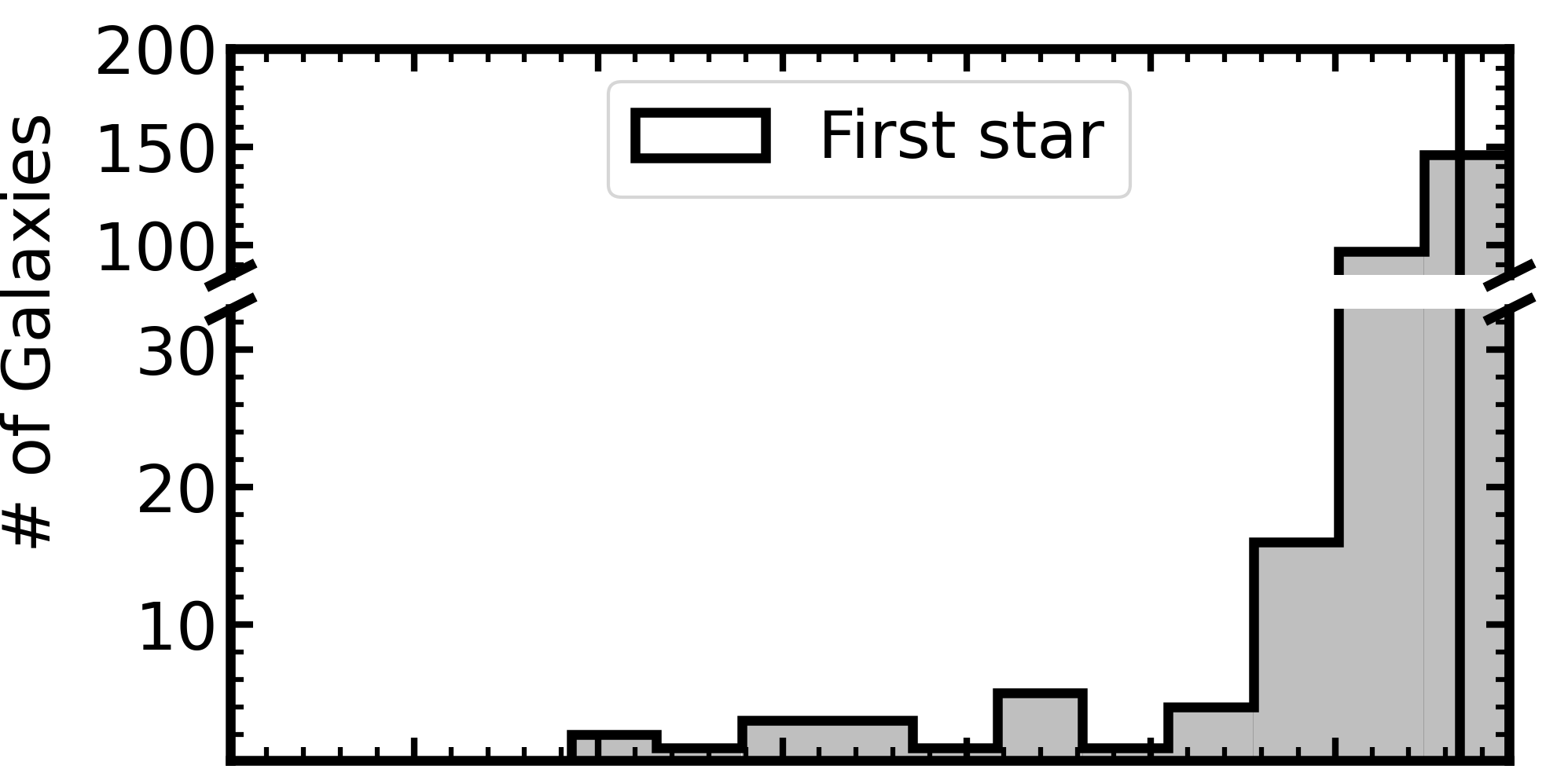}
%    \end{minipage}
%    \begin{minipage}{0.5\textwidth}
    \includegraphics[width=0.48\textwidth]{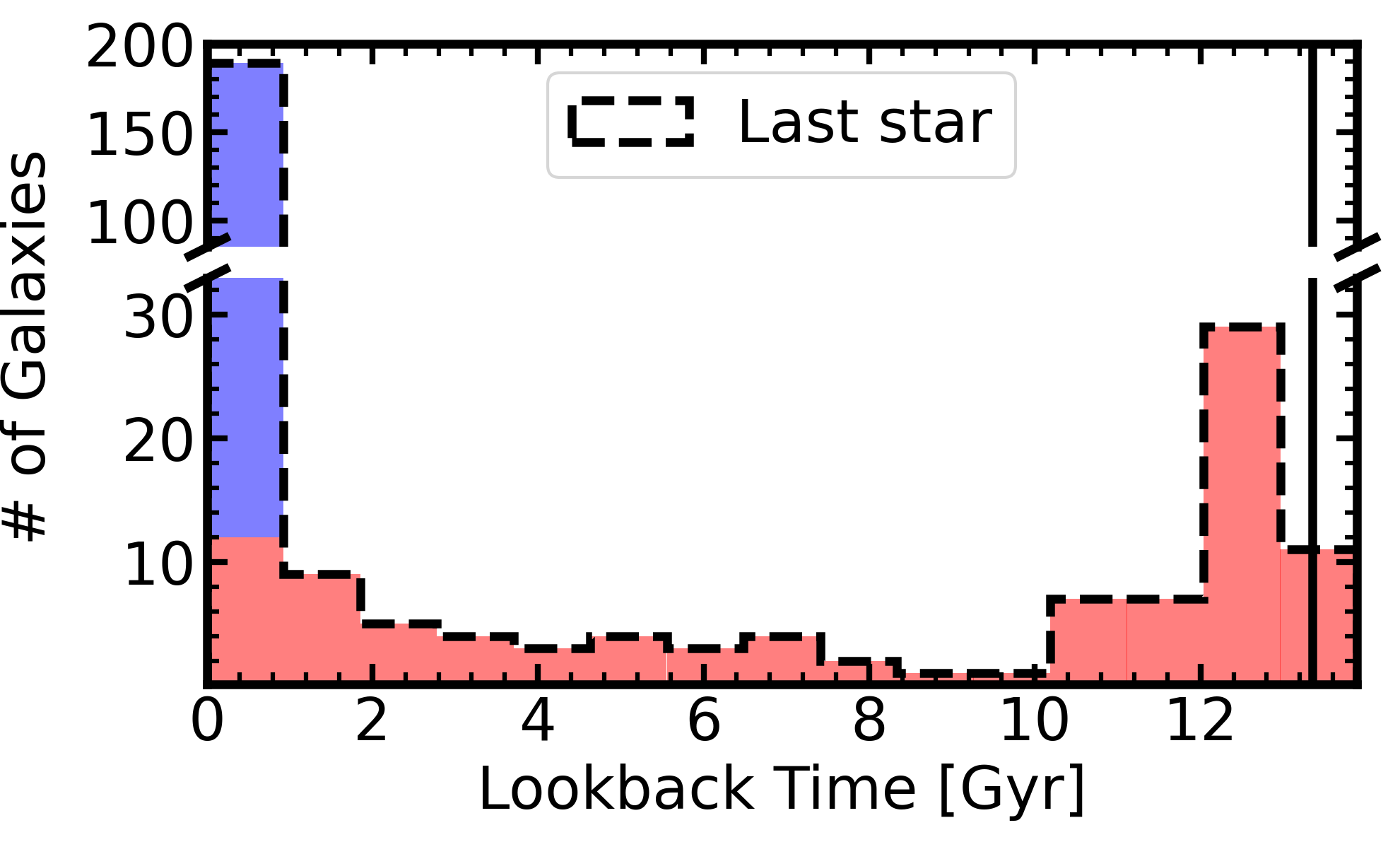}
    \caption{Age distribution of the oldest (top panel) and youngest (bottom) star particles in APOSTLE  field dwarfs identified at $z=0$ (grey data points in Fig.~\ref{fig:mcrit_2}). Most dwarfs start forming stars very early. On the other hand, quiescent galaxies today (red histogram in bottom panel)  ceased forming stars over a wide range of times in the past. Their star formation end times seem to bear no relation with the time of reionization (solid vertical line).  }
    \label{fig:histo_lbt}
\end{figure}

Interestingly as well, there are a few clear outlier halos which start forming stars at masses well below critical. Such outliers tend to have rather low concentrations, opposite to the trend just discussed. We have tracked many of these objects individually, and they correspond to halos whose central gas densities were temporarily enhanced by passage through denser regions of the cosmic web, such as a filament or sheet of gas \citep[see; e.g.,][]{BL13,Wright_2019}. We emphasize, however, that these occurrences appear to be quite rare, as we found only $8$ such systems in the five APOSTLE volumes we have inspected. We intend to study these objects in more detail in future contributions.

Another robust result to glean from Fig.~\ref{fig:mcrit_2} is that few systems form their first stars after $z\sim 2$ or so. Indeed, only $16$ out of $279$ central field dwarfs at the present day started to form stars after $z=2$. This implies that the oldest stars in the majority of $z=0$ luminous systems date to lookback times at least as old as $12$ Gyrs. We show this in the top panel of Fig.~\ref{fig:histo_lbt}, where the grey histogram corresponds to the distribution of ages of the oldest star in all APOSTLE luminous systems with present-day virial masses $M_{200}<10^{11}\, M_\odot$. Note that this is a lower limit to the formation time of the oldest stars, as the artificial PEoS of our simulation is expected to delay the formation of the first stars in many luminous galaxies today. 

It is also clear from the grey circles in Fig.~\ref{fig:mcrit_2} (which correspond to systems identified at $z=0$) that more massive systems tend to have slightly older ancient populations, but the difference in terms of lookback time is small enough for this to be very difficult to discern (recall that the lookback time between $z=2$ and $z=11$ only changes from $10.4$ to $13.3$ Gyrs).

Although the large majority of galaxies begin forming stars very early, there is a discernible population of dwarfs that formed their firsts star more recently, about $4$-$8$ Gyrs ago. This trace population corresponds to dwarfs with uncommon mass accretion histories which first reached the critical mass fairly recently. The origin of these galaxies has been analyzed in detail by~\citetalias{BLF21}, and we refer the interested reader to that work for further details.

Why do the majority of luminous dwarfs in APOSTLE start forming stars early on, regardless of luminosity? The reason may be traced to the mass growth history of individual halos with masses close to the critical mass of $\sim 10^{9.7} M_\odot$ at $z=0$. We show three such mass accretion histories (i.e., the redshift evolution of the mass of the most massive progenitor of a system identified at $z=0$) with blue curves in Fig. ~\ref{fig:mcrit_2}. Two of these, shown in solid blue, correspond to halos hosting luminous galaxies at present, and they start forming stars roughly at the time (identified by open blue circles) that their accretion histories intersect the critical mass curve: systems that cross the critical mass boundary earlier also start forming stars earlier. Indeed, the bottom (dashed) blue curve corresponds to a system that never crosses the critical boundary and that remains ``dark'' at $z=0$. The result of our simulations thus echoes earlier analyses reported by \citet{Fitts_2017}, \citet{Maccio_2017}, and~\citetalias{BLF21}.

The grey histogram in the right-hand panel of  Fig. ~\ref{fig:mcrit_2} shows the $z=0$ halo mass distribution of all {\it luminous} galaxies in APOSTLE. There is a clear peak at the critical mass $\sim 10^{9.7}\, M_\odot$, and a sharp decline towards lower masses. Indeed, basically all halos below $10^9\, M_\odot$ remain ``dark''.  

As discussed by~\citetalias{BLF20}, this decline may be traced to the typical accretion histories of halos in this mass range, which is roughly parallel to the evolution of the critical mass. In other words, most halos that are today above critical have been so since early times, and the same is true for most sub-critical halos. Some halos just below the critical boundary may also harbour luminous systems.  These ``sub-critical'' halos were above critical at some point in their history before their mass growth slowed enough to fall under the critical boundary by the present time. One example of this is shown by the bottom solid blue curve in Fig.~\ref{fig:mcrit_2}.

\subsection{Halo mass growth history and the modulation of star formation}

\begin{figure}
    \centering
    \includegraphics[width=0.49\textwidth]{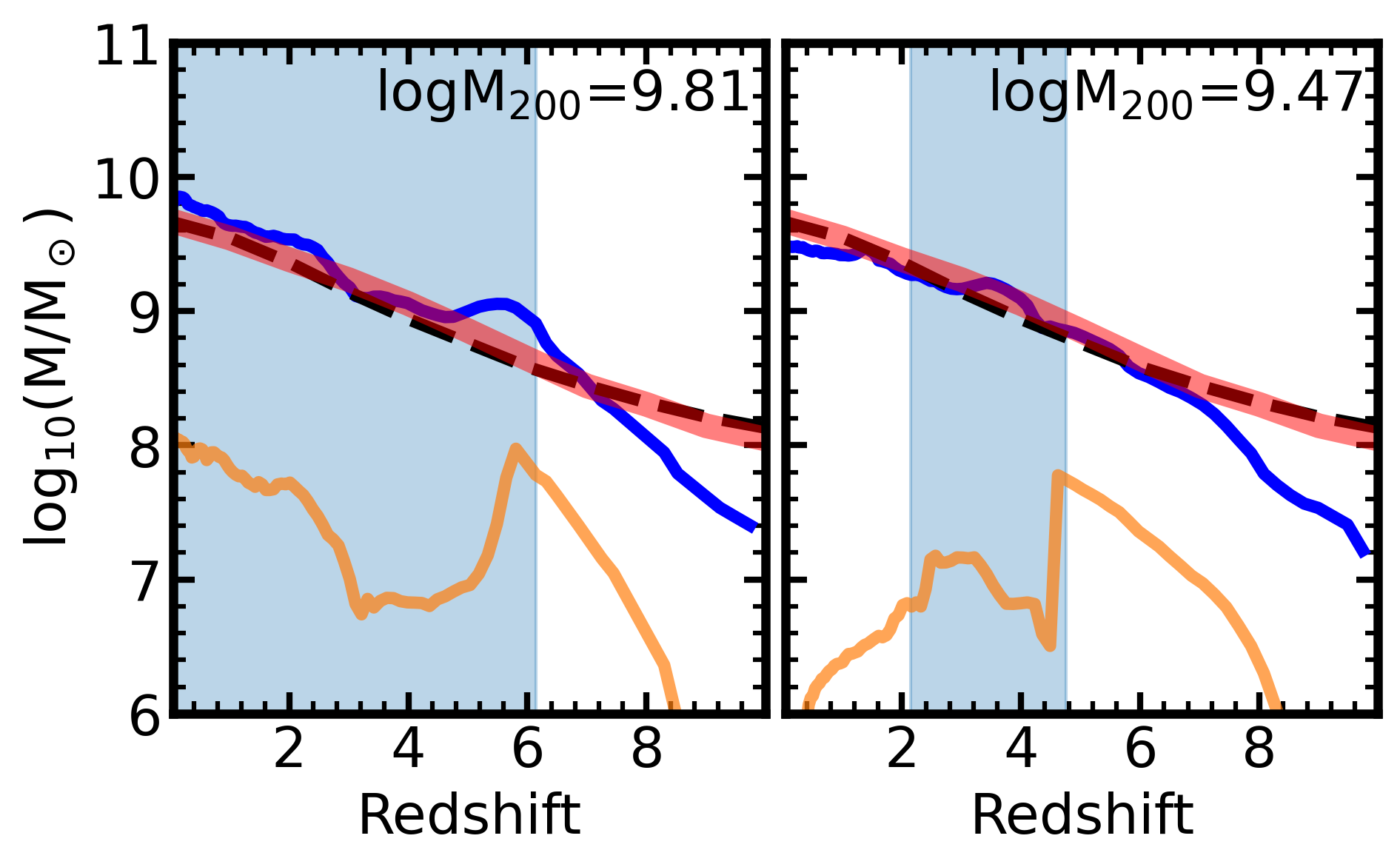}
    \caption{Mass assembly histories of two galaxies in the APOSTLE simulation, with masses at present day near the critical mass. The mass of the most massive progenitor is shown by the blue lines, and its gas mass in orange. The time interval over which a galaxy forms stars is shaded in blue, and is seen to coincide with the time the halo is above the critical boundary, shown by the dashed black curve. The dot-dashed lines show the average mass assembly history of $\Lambda$CDM halos of present-day mass, $M_{\rm crit}$. For reference, the average mass assembly history of halos with present day mass of $10^{9.7}$~ $M_\odot$ is shown by the thick red line.}
    \label{fig:mah}
\end{figure}

The discussion of the previous subsection makes clear that APOSTLE halos begin forming stars only once their mass histories take them into the ``above-critical'' regime. What happens if they happen to fall below critical at later times? We explore this in Fig.~\ref{fig:mah}, where we plot the mass accretion history of two illustrative examples, as well as the average mass accretion history of halos with present day mass equal to the critical mass (thick red line). The blue solid curves indicate the mass evolution of the most massive progenitor, whereas the black dashed curve delineates the critical mass boundary. The orange curves track the gas content of each halo. 

The blue shaded region brackets the interval between the youngest and oldest star formed in each system. The system on the left climbs above critical at $z\sim 7$ and remains so until the present. It starts forming stars soon after becoming critical and is still forming stars at $z=0$. On the other hand, the example on the right depicts a halo that climbs above critical at $z\sim 6$ but becomes sub-critical soon thereafter, at $z\sim 2$. As the shaded region indicates, this halo only forms stars for as long as its mass remains above critical. 

Note that the same process could, in principle, explain why some galaxies stop forming stars for a relatively long time before reigniting, or why some have experienced several distinct episodes of star formation separated by quiescent periods. These cases have been reported in simulations by, e.g., \citet{BL15,BL16,Rey_2020} and \citet{Wright_2019}, although the latter authors interpreted their results in terms of environmental effects, rather than as a result of a critical mass imposed by the ionizing UV background.
%Note, however, that the evolution of the critical mass roughly parallels the average mass assembly history of $\Lambda$CDM halos of the same today (shown by the dot-dashed lines), indicating that the majority of halos above or below critical today have remained so since early times.
We plan to analyze further the relation between episodic star formation and mass accretion histories in a future contribution.
 
The discussion above suggests that the interplay between mass growth history and the critical boundary determines, to a large extent, the star formation history of a dwarf inhabiting a halo of mass comparable to the critical mass. Feedback may also play a role, as seen by the sudden decrease in gas mass (orange line) that accompanies the onset of star formation (Fig.~\ref{fig:mah}), but it appears to be less important overall: despite continuous feedback from ongoing star formation, the system on the left-hand panel of Fig.~\ref{fig:mah} retains some gas until today. This result suggests that the critical mass should roughly delineate a boundary between star-forming and quiescent systems, an issue we examine next.

\subsection{Star-forming vs quiescent dwarfs}

\begin{figure}
    \centering
    \includegraphics[width=0.5\textwidth]{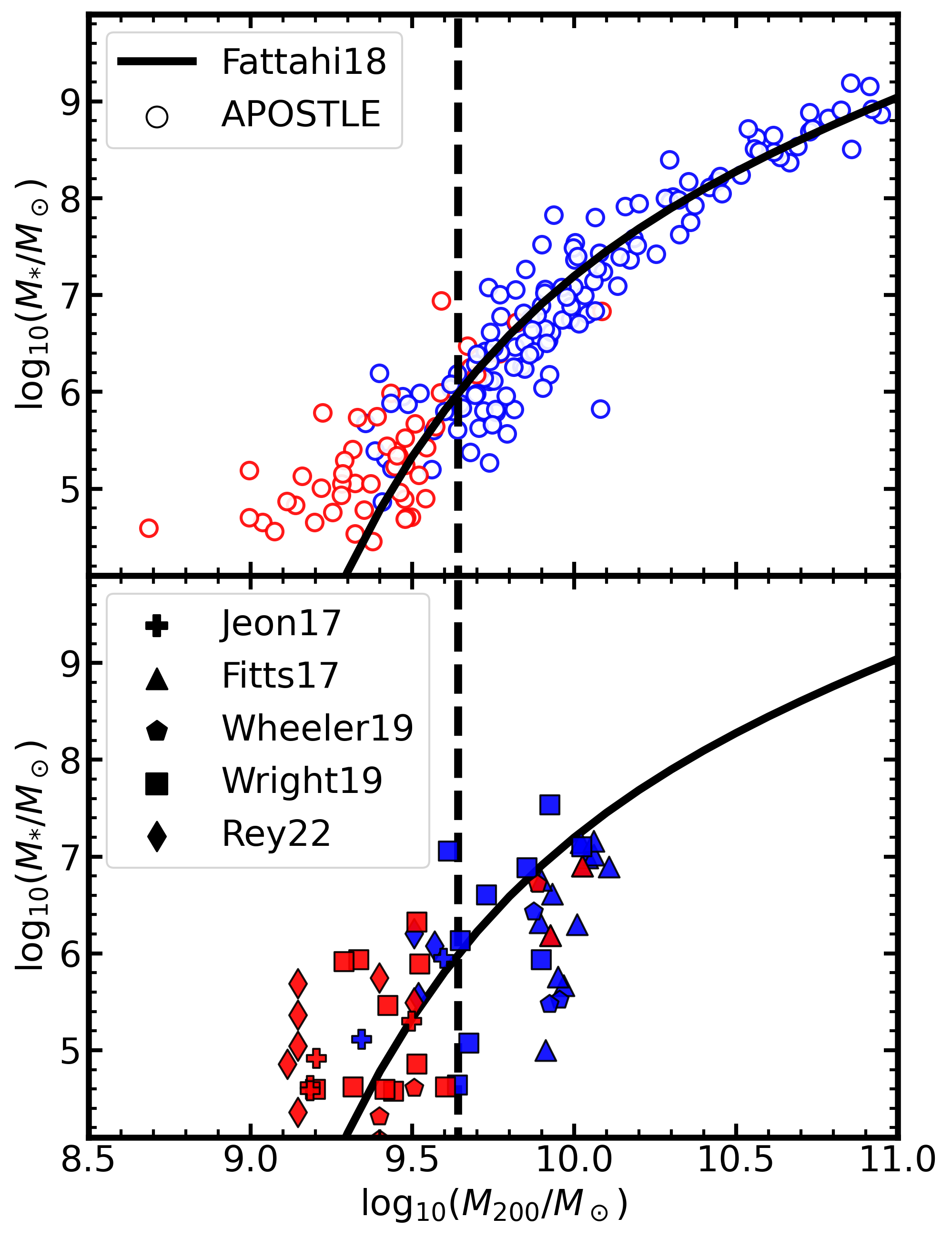}
  \caption{{\it Top:} Stellar mass of star-forming (blue) and quiescent (red) galaxies in APOSTLE, as a function of virial mass at $z=0$. For reference, the solid black curves show the APOSTLE $M_*$-$M_{200}$ relation fit by \citet{Fattahi_2018}. {\it Bottom:} As top panel, but for recent cosmological hydrodynamical simulations of field dwarfs \citep{Jeon_2017,Fitts_2017,Wheeler_2019,Wright_2019,Rey_2022}.}
    \label{fig:ms_v_m200}
\end{figure}

Fig.~\ref{fig:ms_v_m200} examines the star formation properties of APOSTLE dwarfs at $z=0$. The upper panel shows the stellar mass-halo mass relation, coloured blue (star-forming) if star formation is still ongoing and red (quiescent) if no stars have formed in the past $0.5$ Gyrs. The vertical dashed line indicates the critical mass, and clearly separates the two dwarf populations: most halos whose mass is above critical host galaxies where star formation is ongoing, whereas sub-critical halos host almost exclusively quiescent dwarfs. The distinction between these two populations is less clear using stellar mass, although there is a clear trend for the faintest dwarfs to be quiescent.

\begin{figure}
    \centering
    \includegraphics[width=0.5\textwidth]{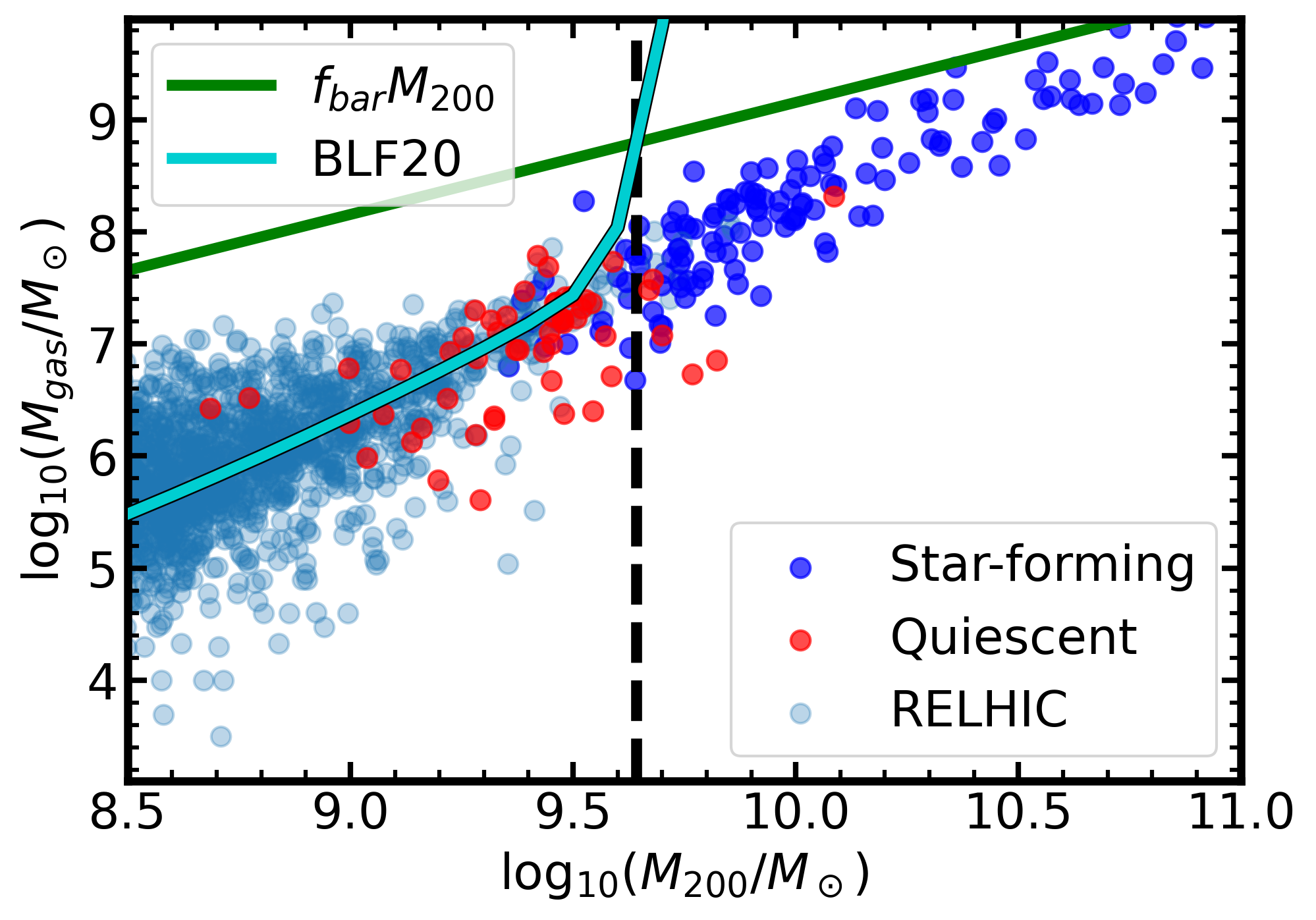}
  \caption{ Gas mass within the virial radius of luminous dwarfs (star-forming in blue;  quiescent in red) and star-less halos (RELHICs, light blue) as a function of $M_{200}$ at $z=0$. %Quiescent galaxies without gas are shown as vertical red lines at $M_{\rm gas}=10^{3.3}\, M_\odot$.
  Masses corresponding to the universal baryon fraction are shown by the green-line. The total gas mass expected for RELHICs from our modeling is  shown by the cyan curve (see also Fig.~\ref{fig:mgas_c}).}
    \label{fig:mg_v_m200}
\end{figure}

There have been a number of suggestions for truncating star formation in field dwarfs, notably the loss of its surrounding gaseous halo due to ram-pressure effects from the cosmic web \citep{BL13} or from potential grazing passages through the virial radius of a more massive system \citep[e.g.,][]{Teyssier_2012}. These may be contrasted with our scenario by examining the gas content of the quiescent population in APOSTLE, which would be largely absent if ram-pressure effects were the dominant mechanism. We explore this in Fig.~\ref{fig:mg_v_m200}, where we plot the gas mass of each APOSTLE central halo at $z=0$ as a function of halo virial mass.

Blue and red circles indicate star-forming and quiescent dwarfs, respectively, while the semi-transparent light-blue symbols indicate the (more numerous) ``dark'' RELHIC systems. The green curve indicates, for reference, the total baryon mass of a halo, $f_{\rm bar} M_{200}$, whereas the cyan curve indicates the total gas mass that results from applying the~\citetalias{BLF20} model and the APOSTLE $T$-$\rho$ relation. As expected, the cyan curve provides a good estimate of the total gas mass of a halo in the sub-critical regime, as already pointed out by~\citetalias{BL17} for RELHICs.

%There are clearly some gas-free (likely ram pressure-stripped) halos for $M_{200}<10^9\, M_\odot$ (i.e., those with $M_{\rm gas}< 10^{4.5}\, M_\odot$, equivalent to one or two gas particles. Galaxies with no gas particles are also shown, as vertical dashes, but very few such systems overlap with the quiescent population (shown in red). 

Note that most quiescent galaxies retain gas, typically one or two orders of magnitude more than the stars they have been able to form. These systems thus appear to be quiescent not because of gas removal by feedback or environmental effects but because their halo masses are below critical and thus unable to compress the gas to high enough densities to ignite star formation.

This constitutes a robust and simple prediction that should be testable by observations. In other words, our simulations suggest the existence of a sizable population of quiescent field dwarfs at the faint end of the luminosity function. Note that this prediction stems from the existence of a minimum halo mass required for the gas to collapse and form stars, and is, therefore, qualitatively independent of the adopted APOSTLE modelling. Such dwarfs are quite rare in the Local Group, with few known examples: the Cetus \citep{Whiting_1999} and Tucana \citep{Lavery_1992} dSphs, and two more distant dwarfs, KKR 25 \citep{Karachentsev_2001,Makarov_2012}, and KKs 3 \citep{Karachentsev_2015}. They also seem to be rare in the local Universe; \citet{Geha_2012} report the existence of a ``threshold of $M_* < 10^9\, M_\odot$ below which quenched galaxies do not exist in the field''. Their survey, however, only extends down to $M_*>10^7\, M_\odot$ before becoming severely incomplete. As shown in Fig.~\ref{fig:ms_v_m200}, the population of quiescent field dwarfs predicted by our analysis is expected to become prominent at much lower stellar masses.

APOSTLE is not the only simulation suite where these two populations of dwarfs have been seen. The bottom panel of Fig.~\ref{fig:ms_v_m200} is analogous to the top, but includes results from five recent cosmological hydrodynamical simulations of dwarf galaxy formation \citep{Jeon_2017,Fitts_2017,Wheeler_2019,Wright_2019, Rey_2022}. Remarkably, although each of these simulations adopts different recipes for star formation/feedback and disparate treatments of the interstellar medium, taken together they seem to agree with our main conclusion: there is a critical halo mass that separates star-forming from quiescent dwarfs. Note again that the separation is much less clear in $M_*$ than it is in virial mass, highlighting the fact that it is the critical mass imposed by the ionizing UV background, and not stellar feedback, the main culprit for the origin of these two populations.

The existence of a population of faint, quiescent dwarfs inhabiting sub-critical halos is thus an intriguing prediction that should be possible to verify with future observations. The quiescent isolated dwarf ($M_* \approx 2\times 10^6\, M_\odot $) recently discovered by \citet{Polzin_2021}, together with the newly identified Tucana B ultra-faint dwarf \citep{Sand_2022}, could very well be the archetypes of a whole population still awaiting discovery.

\subsection{Star formation end times}

As discussed in Sec.~\ref{SecIntro}, cosmic reionization is often assumed to imply a sharp and very early truncation of star formation in faint dwarfs. Although this description may apply to a minority of galaxies inhabiting very low-mass halos well below the critical boundary, we have seen that this is not the case for the majority of field dwarfs inhabiting near-critical halos at $z=0$. In such systems, the ionizing UV background regulates the end of star formation in conjunction with the accretion history of a halo. It is therefore interesting to ask, for dwarfs that have ceased forming stars at present (i.e., those in ``sub-critical'' halos), when they experienced the last episode of star formation.

We show this in the bottom panel of Fig.~\ref{fig:histo_lbt}, where the red and blue histograms delineate the distribution of the youngest star particle in all APOSTLE galaxies with $M_{200}< 10^{11}\, M_\odot$ at $z=0$. Those shaded in blue indicate star-forming systems, whereas those in red correspond to the quiescent population at the present time. Clearly, there is a large diversity of ``quenching'' times, driven by the large diversity of individual halo mass accretion histories at fixed halo mass. This is another robust prediction for the quiescent population of isolated dwarfs that could be addressed in future observational studies.

%Fig.~\ref{fig:tlast_m200} shows the lookback time at which the last star formed in each halo, as a function of present day mass. Intermediate galaxies, with a mass close to the present day critical mass, have a varied distribution of quenching times, indicative of their mass co-evolution with the critical mass. Similarly, galaxies significantly below the critical mass today have been below the critical mass since early times and quenched very soon after forming their stars. 

\subsection{Redshift dependence of the quiescent population}

\begin{figure}
    \centering
    \includegraphics[width=0.45\textwidth]{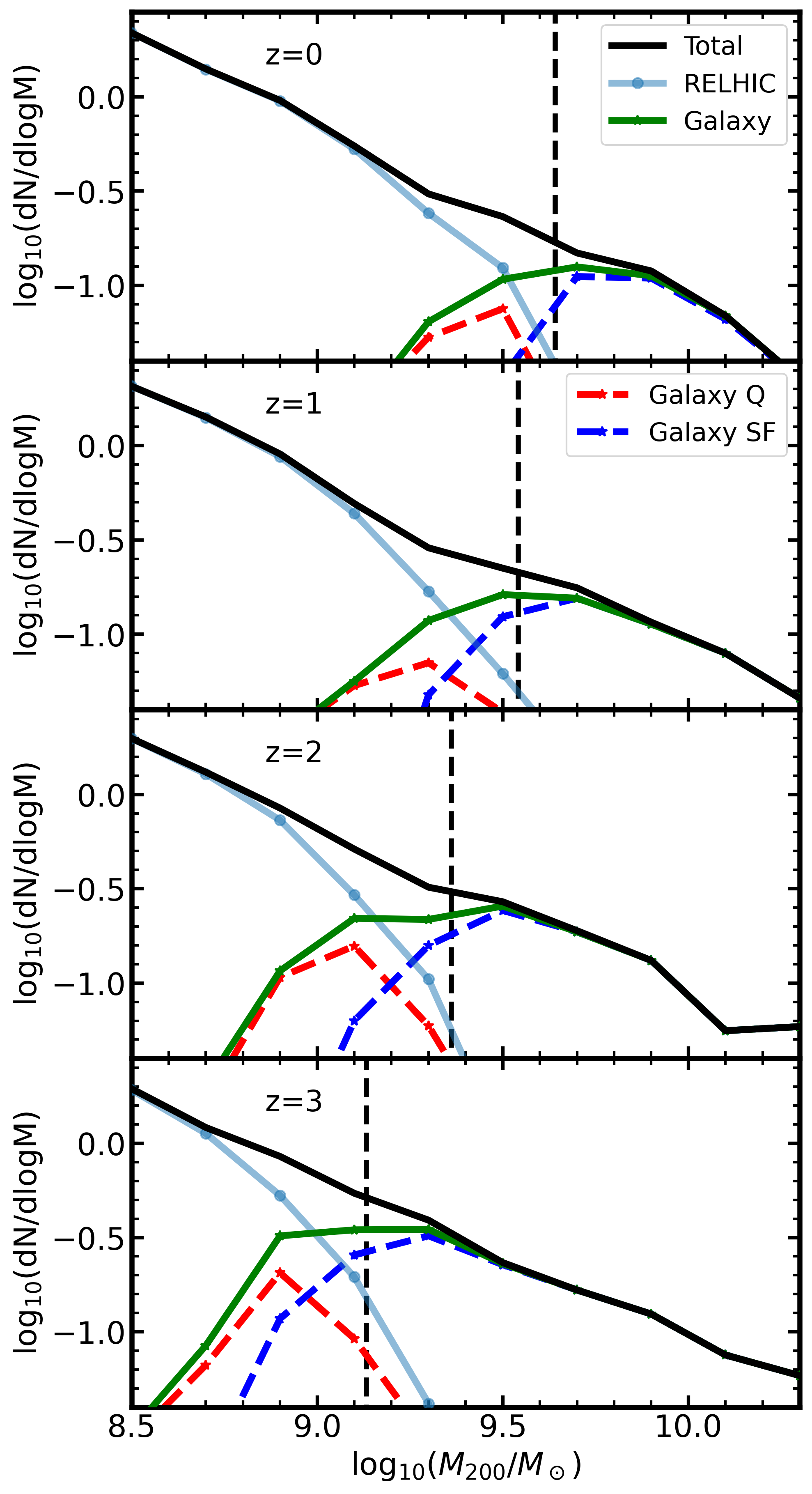}
    \caption{ APOSTLE halo mass functions at different redshifts. Solid black lines indicate the halo mass function, averaged over the five APOSTLE volumes considered in this study. Star-less halos (RELHICs) are shown in light-blue, and galaxies in green. Galaxies are further subdivided into star-forming (blue) and quiescent (red), with the critical mass at each redshift shown as a vertical dashed line.}
    \label{fig:hmfs}
\end{figure}

Our discussion so far implies that the quiescent population of dwarfs we discussed above should exist at all redshifts, although the boundary between quiescent and star-forming should shift to lower masses at increasing redshift, tracking the evolution of the critical mass. We explore this in Fig.~\ref{fig:hmfs}, where we show the differential halo mass function (averaged over the five APOSTLE volumes) at different redshifts. The solid black lines indicate the dark halo mass function, and the cyan curve denotes halos that have remained dark at each redshift. Green corresponds to all luminous galaxies, split between star-forming (blue) and quiescent (i.e., those that did not form any stars in the most recent $0.5$ Gyrs, in red) populations. The vertical dashed line indicates the APOSTLE critical mass (black curve in Fig.~\ref{fig:mcrit}).

The quiescent population is mostly contained below the critical mass threshold at all redshifts, indicating that the critical mass model is still a valid threshold for star formation at other redshifts. The differentiation between populations becomes less neat at higher redshift, with an increasing function of sub-critical halos hosting star-forming dwarfs. This is most likely because our definition of ``star-forming'' uses a fixed time window of $0.5$ Gyr to categorize systems, which represents a significant fraction of the universe's age at earlier times. 

Another result illustrated by Fig.~\ref{fig:hmfs} is that, at all redshifts, $M_{\rm crit}$ represents a characteristic ``threshold'' for galaxy formation, in the sense that the fraction of ``dark'' halos grows sharply below that mass. Indeed, the number density of galaxies peaks at about the critical mass so that, in terms of sheer numbers, the majority of field galaxies in any given volume are faint dwarfs inhabiting halos whose mass is near the critical boundary.

\section{Summary and conclusions}
\label{SecConc}

We have examined the onset of star formation in low-mass halos identified in the APOSTLE cosmological hydrodynamical simulations. In agreement with the work of~\citetalias{BLF20}, we find that star formation begins once the mass of a halo reaches a characteristic ``critical'' mass that may be derived using a simple model that combines the NFW mass profile of $\Lambda$CDM halos and the thermodynamics of gas heated by the cosmic ionizing UV background.

Our modelling assumes hydrostatic equilibrium to derive the gas density profile at various redshifts and identifies the critical mass where the central gas density equals the gas density threshold for star formation adopted in EAGLE/APOSTLE. The critical mass defined in this way agrees well, post reionization, with the critical mass derived by~\citetalias{BLF20}. This redshift-dependent critical mass describes quite accurately the minimum virial mass needed for APOSTLE halos to first form stars, especially after reionization.

According to this model, star formation should cease (or never start) in sub-critical halos, in excellent agreement with APOSTLE simulation results and other recent cosmological hydrodynamical simulations. The ionizing UV background thus seems to be the main mechanism regulating star formation in dwarf galaxies through the interplay between the critical mass boundary and the mass accretion history of a dwarf's dark halo. 

In addition, the evolution of the critical mass with time is roughly parallel to the average mass accretion history of halos near the critical regime, implying that, in general, halos with mass above or below critical at present have remained so since early times. As discussed by~\citetalias{BLF20}, this implies that the critical mass at $z=0$ represents a ``threshold'' below which the fraction of halos harbouring luminous systems drops sharply.

For the same reason, most dwarf galaxies at $z=0$, regardless of their luminosity, started forming stars quite early (lookback time $>12 \rm \ Gyr$), providing a simple and appealing explanation for the ubiquitous presence of ancient stellar populations in all known dwarfs. This is a robust result, independent of the modelling details in our simulations.

Dwarf galaxies inhabiting sub-critical halos today are expected to be quiescent, making up a sizable population of non-star-forming dwarfs at the faint end of the field galaxy stellar mass function. Only a few such galaxies have been discovered in the field so far, but the discovery of this population would provide strong evidence in support of this scenario.

Finally, we speculate that halos whose accretion histories cross the critical boundary several times during their evolution may host several distinct episodes of star formation without the need for environmental effects. This may help to explain the puzzlingly episodic nature of star formation observed in some dwarfs. As most halos above or below critical today have remained so since early times, this is likely to affect only a small fraction of dwarfs.

Although this simple scenario accounts for the main features of the star formation history of the faintest dwarfs in APOSTLE, it is important to note some of its caveats and limitations. We have focused exclusively on isolated systems, mainly because of simplicity, but note that environmental effects such as ram-pressure and tidal stripping may dominate in dwarfs that are satellites of more massive systems. 

Therefore, one should exercise care when applying these results to interpret the star formation histories of dwarfs in the Local Group, where satellites currently make up the majority of systems studied observationally in detail.

We also note that using a ``polytropic equation of state'' (PEoS) in EAGLE/APOSTLE artificially reduces the ability of low-mass halos to form stars prior to reionization. This rather crude numerical treatment of high-density gas means that our conclusions, however appealing, must be treated with care and should be scrutinized further in future simulation work with more sophisticated treatments of the interstellar medium. Nevertheless, we believe that many of the conclusions highlighted above will prove of lasting value and will provide a useful interpretive framework for future work.

\section*{Acknowledgements}
We wish to acknowledge the generous contributions of all those who made possible the Virgo Consortium’s EAGLE/APOSTLE simulation projects.  This work used the DiRAC@Durham facility managed by the Institute for Computational Cosmology on behalf of the STFC DiRAC HPC Facility (www.dirac.ac.uk). The equipment was funded by BEIS capital funding via STFC capital grants ST/K00042X/1, ST/P002293/1, ST/R002371/1, and ST/S002502/1, Durham University and STFC operations’ grant ST/R000832/1. DiRAC is part of the National e-Infrastructure. MPW acknowledges receipt of an NSERC-CGSM and of an R. M. Petrie Memorial Fellowship in support of this work.  ABL acknowledges European Research Council (ERC) under the European Union’s Horizon 2020 research and innovation program under the grant agreements 101026328 and 757535, and UNIMIB’s Fondo di Ateneo Quota Competitiva (project 2020-ATESP-0133).
\section*{Data Availability}
The simulation data used in this article can be shared on reasonable
request to the corresponding author.

%\pagebreak
\bibliographystyle{mnras}
\bibliography{main}

\end{document}